\documentclass[acmsmall]{acmart}

\usepackage[stable]{footmisc}
\usepackage{amsmath}
\usepackage{mathtools}
\usepackage{algorithm}
\usepackage{algpseudocode}
\usepackage[group-separator={,}]{siunitx}
\usepackage{xspace}
\usepackage{cleveref}
\usepackage{booktabs}
\usepackage{tabu}
\usepackage{tabularx}
\usepackage{tabularray}
\AtBeginDocument{\hypersetup{pdfborder = {0 0 0}}}
\usepackage{graphbox}
\usepackage{diagbox}
\usepackage{threeparttable}
\usepackage{stfloats}
\usepackage{wrapfig}
\usepackage{stmaryrd}
\usepackage{caption}
\usepackage{subcaption}
\usepackage{bbding}
\usepackage{multirow}
\usepackage{tikz}
\usepackage{pifont}
\usepackage{pgfkeys}
\usepackage{enumitem}
\usepackage[framemethod=tikz]{mdframed}
\usepackage{extarrows}
\usetikzlibrary{decorations.pathreplacing}
\usetikzlibrary{shapes.multipart}
\usetikzlibrary{graphs}
\usetikzlibrary{graphs.standard}

\usepackage{array}
\usepackage[export]{adjustbox}

\Crefname{figure}{Fig.}{Fig.}

\usepackage{rotating}

\usepackage[frozencache=true,cachedir=minted-cache]{minted}
\setminted{
    framesep=0pt,
    fontsize=\footnotesize
}

\makeatletter
\newenvironment{tabminted}{%
  \let\FV@ListVSpace\relax  
  \minted
}{%
  \endminted
  \unskip   
  \aftergroup\@tabmintedend
}
\newcommand*{\tabminted@finalstrut}[1]{%
  \ifdim\prevdepth>0pt
    \ifdim\dp#1>\prevdepth
      \vskip\dimexpr(\dp#1)-\prevdepth\relax
    \fi
  \else
    \vskip\dimexpr(\dp#1)\relax
  \fi
}
\newcommand*{\@tabmintedend}{%
  \let\@finalstrut\tabminted@finalstrut
}
\makeatother

\setcopyright{rightsretained}
\acmPrice{}
\acmDOI{10.1145/3622822}
\acmYear{2023}
\copyrightyear{2023}
\acmSubmissionID{oopslab23main-p227-p}
\acmJournal{PACMPL}
\acmVolume{7}
\acmNumber{OOPSLA2}
\acmArticle{246}
\acmMonth{10}
\received{2023-04-14}
\received[accepted]{2023-08-27}

\begin{document}

\title{Mat2Stencil: A Modular Matrix-Based DSL for Explicit and Implicit Matrix-Free PDE Solvers on Structured Grid}

\author{Huanqi Cao}
\orcid{0000-0002-3870-106X}
\email{caohq18@mails.tsinghua.edu.cn}
\affiliation{%
  \institution{Tsinghua University}
  \city{Beijing}
  \country{China}
}

\author{Shizhi Tang}
\orcid{0000-0002-6543-0859}
\email{tsz19@mails.tsinghua.edu.cn}
\affiliation{%
  \institution{Tsinghua University}
  \city{Beijing}
  \country{China}
}

\author{Qianchao Zhu}
\orcid{0009-0001-5021-2912}
\email{dysania@pku.edu.cn}
\affiliation{%
  \institution{Peking University}
  \city{Beijing}
  \country{China}
}

\author{Bowen Yu}
\orcid{0000-0001-5537-8244}
\email{yubowen@tsinghua.edu.cn}
\affiliation{%
  \institution{Tsinghua University}
  \city{Beijing}
  \country{China}
}

\author{Wenguang Chen}
\orcid{0000-0002-4281-1018}
\email{cwg@tsinghua.edu.cn}
\affiliation{%
  \institution{Tsinghua University}
  \city{Beijing}
  \country{China}
}
\affiliation{%
  \institution{Pengcheng Laboratory}
  \city{Shenzhen}
  \country{China}
}

\newcommand{\SYS}{\textsc{Mat2Stencil}\xspace}
\newcommand{\SYSfull}{\SYS}
\newcommand{\FT}{FreeTensor\xspace}
\newcommand{\bcao}[1]{{\textcolor{red}{[BCao: #1]}}}
\newcommand{\YBW}[1]{{\textcolor{purple}{[YBW: #1]}}}
\newcommand{\tsz}[1]{{\textcolor{blue}{[TSZ: #1]}}}

\setcounter{MaxMatrixCols}{20}

\newenvironment{fbmatrix}{\footnotesize\begin{bmatrix}}{\end{bmatrix}\ignorespacesafterend}

\newcommand*\Let[2]{\State #1 $\gets$ #2}


\begin{abstract}

Partial differential equation (PDE) solvers are extensively utilized across numerous scientific and engineering fields. However, achieving high performance and scalability often necessitates intricate and low-level programming, particularly when leveraging deterministic sparsity patterns in structured grids.

In this paper, we propose an innovative domain-specific language (DSL), Mat2Stencil, with its compiler, for PDE solvers on structured grids.
Mat2Stencil introduces a structured sparse matrix abstraction, facilitating modular, flexible, and easy-to-use expression of solvers across a broad spectrum, encompassing components such as Jacobi or Gauss-Seidel preconditioners, incomplete LU or Cholesky decompositions, and multigrid methods built upon them.
Our DSL compiler subsequently generates matrix-free code consisting of generalized stencils through multi-stage programming. The code allows spatial loop-carried dependence in the form of quasi-affine loops, in addition to the Jacobi-style stencil's embarrassingly parallel on spatial dimensions.
We further propose a novel automatic parallelization technique for the spatially dependent loops, which offers a compile-time deterministic task partitioning for threading, calculates necessary inter-thread synchronization automatically, and generates an efficient multi-threaded implementation with fine-grained synchronization.

Implementing 4 benchmarking programs, 3 of them being the pseudo-applications in NAS Parallel Benchmarks with $6.3\%$ lines of code and 1 being matrix-free High Performance Conjugate Gradients with $16.4\%$ lines of code, we achieve up to $1.67\times$ and on average $1.03\times$ performance compared to manual implementations.
\end{abstract}



\maketitle

\section{Introduction}

\footnote{We improved the writing of this section by prompting GPT-4 \cite{openai2023gpt4}, OpenAI’s large-scale language-generation model, with the draft written by humans. Upon generating, we reviewed, edited, and revised the language to ensure its fidelity and take ultimate responsibility for the content of this publication.}
Numerical solutions of partial differential equations (PDEs) are crucial in a wide variety of scientific computing applications, spanning from weather prediction \cite{wrf} to seismic imaging \cite{seismic}.
Many of these applications employ structured grids, primarily Cartesian grids, for spatial discretization.
These grids manifest as 2-dimensional or 3-dimensional arrays, with calculations executed through algorithm-driven traversal over the array elements.
In the past, scientists employed general-purpose programming languages, such as FORTRAN and C/C++, to manually develop solvers for PDEs on structured grids.
This process required extensive use of ``for'' loops and arithmetic operations, often leading to solvers encompassing thousands of lines of code.
As multi-threading and single instruction multiple data (SIMD) technologies have advanced in modern computing, the manual development of such solvers is progressively becoming less practical and less effective.

\begin{figure}[h]
    \centering
    \begin{subfigure}[t]{0.45\textwidth}
        \centering
        \includegraphics[width=0.8\textwidth]{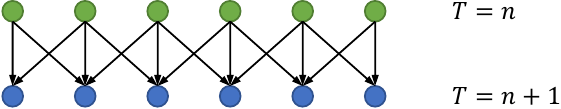}
        \subcaption{The dependence graph in an explicit solver.}
        \label{fig:intro-example:explicit}
    \end{subfigure}
    \hfill
    \begin{subfigure}[t]{0.45\textwidth}
        \centering
        \includegraphics[width=0.8\textwidth]{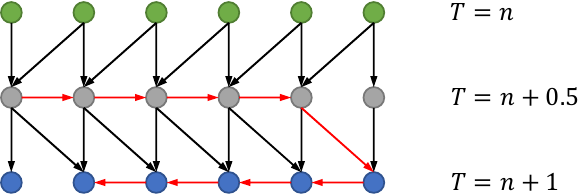}
        \subcaption{The dependence graph in an implicit solver, demonstrating a symmetric Gauss-Seidel iteration. Critical path marked red.}
        \label{fig:intro-example:implicit}
    \end{subfigure}
    \vspace{-1em}
    \caption{The different computation patterns of explicit and implicit solving algorithms, with a 1D heat equation (\Cref{eq:heat-eq}) as an example.}
    \vspace{-1em}
    \label{fig:intro-example}
\end{figure}

A natural direction to address this issue is to translate the operations on structured grids to manipulations of sparse matrices, which we term the \textit{matrix-based} approach.
The PDE solving procedure thus becomes multiplying sparse matrices to vectors (\texttt{SpMV}) and solving sparse linear systems according to the matrix.
This approach thus typically employs general library routines \cite{intel-mkl,scipy} to work on the user-provided matrices.
However, there are two critical issues.
First, constructing the sparse matrix for the PDE problem necessitates significant human effort due to the complexity of real-world equations and their solvers.
Second, as these libraries serve general purposes, they must read the matrix and analyze it at runtime to cater to a broader range of applications, including unstructured grids.
Consequently, they have limited ability to efficiently exploit structured sparsity patterns.
Despite covering the domain well, the matrix-based approach suffers from unresolved programmability and performance issues.

However, \textit{matrix-free} solvers that directly target the equations are typically handwritten in languages such as C/C++ or FORTRAN, making them labor-intensive to implement.
Domain-specific languages (DSLs) and compilers have emerged as promising approaches to simplify programming and enhance performance.
By eliminating matrices, the computations involved in solving partial differential equations (PDEs) are reduced to stencil operators, which iterate over grid points and compute new values based on their neighborhoods.
A well-known class of stencil operators is the Jacobi-style stencils, which read and write different arrays at each time step and exhibit no spatial dependence, which is the dependence between different spatial locations at the same time step, as illustrated in \Cref{fig:intro-example:explicit}.
In contrast, Seidel-style stencils read and write the same arrays, thereby introducing spatial dependence, as seen in \Cref{fig:intro-example:implicit}, which makes them challenging to automatically parallelize and optimize.
While some DSLs, such as \cite{devito,exastencil}, have successfully simplified the programming of PDE equations and solvers by primarily targeting Jacobi-style stencils, they fail on many algorithms containing Seidel-style stencils due to limited abstractions.
On the other hand, domain-specific compilers, such as \cite{ppcg,pluto,diamond}, support both types of stencil operators through advanced dependence analysis techniques but lack abstractions for PDEs, necessitating significant human effort to program a solver with nested loops.

In summary, we recognize several challenges in structured grid PDE solving:
\ding{182} a labor-intensive process to express the equation and the solver;
\ding{183} a conflict between expressiveness and the matrix-free property;
\ding{184} the absence of effective automatic optimization for Seidel-style stencils.
To address these challenges, we introduce a novel DSL called \SYS, designed for both explicit and implicit solving algorithms for PDEs, while ensuring its expressiveness, modularity, and performance.
Our contributions are as follows:

\begin{itemize}
    \item We enable modular programming of equations and their solvers through a high-level abstraction of \textit{structured sparse matrices}.
    It utilizes a set of simple matrices and their arithmetic operations to compose more complicated problem matrices.
    These pre-defined matrices reflect the various steps involved in discretizing partial differential equations (PDEs) and enable user-friendly problem construction in PDE-solving processes.
    \item We enhance expressiveness by introducing a low-level abstraction for structured sparse matrices based on \textit{row functions}.
    It allows users to effortlessly define their custom sparse matrices as well as implement custom algorithms against any structured sparse matrices.
    The design is kept simple and compile-time resolvable, to enable further lowering to matrix-free implementations.
    \item We compile structured sparse matrices user code to matrix-free target code by expanding the sparse matrices during the first stage of our multi-stage programming infrastructure.
    In specific, we backtrack and discuss different cases of loop indices to fill the gap between compile-time (stage 1) and runtime (stage 2) control flow, generating matrix-free code consisting of regular quasi-affine loops.
    \item We develop a novel automatic parallelization technique, SewTh (abbrev. for \textbf{Sew}ing the \textbf{Th}reads), based on polyhedral dependence analysis, to enable efficient multi-threaded execution of the more generalized stencil loops with spatial dependencies. This approach partitions the loop domain across threads and automatically generates fine-grained synchronization when dependencies cross thread boundaries.
\end{itemize}

Altogether, our new DSL is capable of expressing a much wider range of explicit and implicit PDE solvers.
We evaluate our DSL by implementing four benchmarking programs from the NAS Parallel Benchmarks (NPB) \cite{npb} and High Performance Conjugate Gradients (HPCG) \cite{hpcg}, which includes Symmetric Gauss-Seidel, Symmetric Successive Over-Relax-ation, and direct solving of multi-diagonal matrices (degenerated from Incomplete Lower-Upper decomposition) as the implicit solving kernels with spatial dependence.
Our performance reaches up to $1.67\times$ performance in NPB compared with the original FORTRAN implementation, with only $6.3\%$ lines of code; we also finish HPCG implementation in $16.4\%$ lines of code compared to a manually matrix-free implementation.
The geometric mean of our performance over all the programs is $1.03\times$ of the highly-optimized matrix-free human implementations.

\section{Background and Motivation}\label{sec:bg}

In this section, we discuss the current status and problems in differential equation solving, and then the background of techniques we will employ in this paper, including multi-stage programming and polyhedral analysis for automatic parallelization.

\subsection{Solving Differential Equations on Structured Grids}

In this subsection, we demonstrate existing approaches to solving differential equations on structured grids.
We use the simple equation in \Cref{eq:heat-eq} as an example in which we want to solve $u(x, t)$.
\begin{equation}
    \label{eq:heat-eq}
    \frac{\partial u}{\partial t} = \frac{\partial^2 u}{\partial x^2}, x \in [0, 1]; u(0, t) \text{ and } u(1, t) \text{ constant}.
\end{equation}

To solve it numerically through Finite Difference \cite{finite-diff}, the variable $t$ is discretized temporally to $T \Delta t$ and variable $x$ spatially to $i \Delta x$ ($0 \leq i \leq n, \Delta x = 1/n$), in which $T$ for timesteps and $i$ for grid point number are both integers.
We thus note $u(T \Delta t, i \Delta x)$ as $u^{(T)}_i$ in short.
If we discretize $\partial u/\partial t$ as $(u^{(T+1)}_i - u^{(T)}_i) / \Delta t$, we have the explicit numerical scheme in \Cref{eq:heat-eq-explicit}, which is a iterative approach to compute $u$ one by each timestep:
\begin{align}\label{eq:heat-eq-explicit}
    \nonumber \frac{u^{(T+1)}_i - u^{(T)}_i}{\Delta t} &=
    \begin{cases}
    \left(u^{(T)}_{i-1} - 2u^{(T)}_i + u^{(T)}_{i+1}\right)/(\Delta x)^2 & \text{if } 0 < i < n, \\
    0 & \text{otherwise}
    \end{cases} \\
    \allowdisplaybreaks
    u^{(T+1)}_i &=
    \begin{cases}
    \lambda u^{(T)}_{i-1} + (1 - 2\lambda) u^{(T)}_i + \lambda u^{(T)}_{i+1} & \text{if } 0 < i < n, \\
    u^{(T)}_i & \text{otherwise}
    \end{cases}\ \ \ \text{where } \lambda = \Delta t / (\Delta x)^2.
\end{align}

However, such explicit schemes sometimes suffer from stability problems. Instead, if we discretize $\partial u/\partial t$ as $(u^{(T)}_i - u^{(T-1)}_i) / \Delta t$, we then derive an implicit numerical scheme in \Cref{eq:heat-eq-implicit}, which requires solving a linear system at each timestep:

\begin{align}\label{eq:heat-eq-implicit}
    \nonumber & \frac{u^{(T)}_i - u^{(T-1)}_i}{\Delta t} =
    \begin{cases}
    \left(u^{(T)}_{i-1} - 2u^{(T)}_i + u^{(T)}_{i+1}\right)/(\Delta x)^2 & \text{if } 0 < i < n \\
    0 & \text{otherwise}
    \end{cases} \\
    & \begin{cases}
    -\lambda u^{(T)}_{i-1} + (1 + 2\lambda) u^{(T)}_i - \lambda u^{(T)}_{i+1} = u^{(T-1)}_i & \text{for } 0 < i < n, \\
    u^{(T)}_0 = u^{(T-1)}_0, u^{(T)}_n = u^{(T-1)}_n
    \end{cases}
    \ \ \ \text{where } \lambda = \Delta t / (\Delta x)^2.
\end{align}

We see that explicitly solving differential equations only requires straightforward computation, while an implicit method will need to solve systems of linear equations.
To solve the sparse linear systems, direct methods (usually for multi-diagonal matrices) and iterative methods using Jacobi, Gauss-Seidel, Incomplete LU decomposition, etc., have been employed.
Though in different patterns, the existing methods for both cases can be classified into two: one is matrix-based, generates sparse matrices, and uses general routines for sparse linear algebra; the other is matrix-free and targets the math formulas directly.

\subsubsection{Matrix-based Approach: General Sparse Linear Algebra}\label{sec:bg:sparse}

One general method to deal with the grids that PDEs are discretized onto involves sparse matrices.
Regarding \Cref{eq:heat-eq-implicit}, the linear system to be solved can be formulated in matrix form as follows:
\begin{equation}
    \label{eq:heat-eq-implicit-mat}
    \begin{fbmatrix}
    1 \\
    -\lambda & 1 + 2\lambda & -\lambda \\
    & \ddots & \ddots & \ddots \\
    && -\lambda & 1 + 2\lambda & -\lambda \\
    &&&& 1 \\
    \end{fbmatrix}
    \mathbf{u}^{(T)} = \mathbf{u}^{(T-1)}
\end{equation}

Users may thus generate the sparse matrix by themselves and use general sparse linear algebra routines to solve their equation.
Similarly, the explicit form in \Cref{eq:heat-eq-explicit} is formulated as a similar sparse matrix-vector multiplication, or \textit{SpMV}.
Thus, by implementing those algorithms against arbitrary sparse matrices, library providers liberate domain experts from writing those complicated codes, only leaving the generating of sparse matrices.

Yet, generating the sparse matrix via hand-written codes can be annoying.
With much more complex equations, the sparse matrix may require more than one page of math formulas to specify, not to mention the code.
Besides the difficulty in constructing the matrices, the performance of the matrix-based approach is unsatisfying.
Although general sparse routines are highly optimized, require extensive human efforts from expert programmers, and usually cannot be implemented efficiently by science domain researchers, these general sparse routines cannot compete with matrix-free approaches in structured problems, due to the lack of structural information at compile time.

\subsubsection{Matrix-free Approaches: Stencil by Hand or Automatically}

Regarding the structured grids, it is possible to get rid of the matrices and perform computation directly, which we call ``matrix-free''.
Like in \Cref{eq:heat-eq-explicit}, explicit approaches primarily involve embarrassingly parallel stencil computations in a single timestep.
In each iteration, they read neighbor points in the input 2D or 3D tensors with a certain pattern, go through a certain function as its computation, and yield the corresponding value in output tensor(s).
There is no data dependence between different grid points if we look at only one iteration.
Due to its nature of independent computation, parallelizing such types of stencil computation is straightforward.
We denote such stencils as ``Jacobi-style'' stencils considering that Jacobi iterations match such a computational pattern without purely spatial dependence.
Since the computation is highly customized across applications due to different PDEs to solve, and optimization by hand requires massive human effort, many domain-specific languages (DSLs) have been designed to address the optimization of stencil computation, as well as programmability of explicit or Jacobi iteration-based implicit solvers.

While Jacobi-style stencil computations are widely researched, many implicit solvers require operators not covered by such DSLs.
Such operators, while still operating neighbors on structured grids, have dependence carried purely by spatial loops on one or more dimensions, which we recognize as Seidel-style stencils, named after the most commonly used Gauss-Seidel iteration but not limited to that particular algorithm.
For example, directly solving \Cref{eq:heat-eq-implicit} requires Gaussian eliminating forward and backward, introducing sequential dependence along the $x$ axis.
More complicated equations and numerical schemes may require iterative methods such as multi-grid solvers or conjugate gradients, which in turn requires routines like Gauss-Seidel iteration, incomplete lower-upper (ILU) factorization, fine-grained parallel ILU \cite{fgpilu}, and so on.

Although the 1D case introduced here has to be executed sequentially, Seidel-style stencils with two or more dimensions are usually possible to get parallelized, but not as trivial as those Jacobi-style stencils, requiring more human efforts to optimize.
For example, the block tri-diagonal pseudo-application in NAS Parallel Benchmarks (NPB-BT) is written in over 400 lines of Fortran code for solving the block tri-diagonal matrix in only one direction, and it has three directions to solve, involving over 1300 LOC.
However, such routines are not yet supported by the DSLs targeting PDE solving.
As the best among existing, ExaStencils is capable of implicitly solving differential equations, but only through Jacobian and forward Gauss-Seidel iterations in a multi-grid (MG) solver, which is just a particular case with limited computation patterns.

\subsubsection{Discussion}
Due to the difficulty of manually programming Seidel-style stencil operators, the matrix-based representation is usually preferred, especially in the case of implicit solvers.
For example, the High Performance Conjugate Gradients (HPCG) Benchmark \cite{hpcg}, a widely recognized effort for ranking top-level supercomputers, enforces sparse matrices and forbids matrix-free approaches in its valid implementations, indicating their concern about the generality of operator implementations.
However, it turns out such prohibition limits performance drastically.
It is reported that implementing the HPCG computation in a matrix-free form significantly improves the performance, by $4.67\times$ on the New Sunway supercomputer \cite{newsw-hpcg}.
When possible, HPC researchers still seek matrix-free approaches even for implicit approaches, e.g., the 2016 Gordon Bell Prize winner \cite{gb16} designed and manually implemented a geometry-based ILU that maps the dependence to hardware-supported inter-core communication.

To this end, we seek a solution so that:
\ding{182} the Seidel-style operators are naturally enabled, resolving the coverage issue of stencil DSLs;
\ding{183} the construction of linear problems is easy for users, resolving the programmability issue of the matrix-based approach;
\ding{184} the matrix-free property is retained at runtime, resolving the performance issue of the matrix-based approach.
These challenges lead us to a new DSL working with sparse matrices but compiled into generalized stencils.

\subsection{Multi-stage Programming}

Multi-stage programming, or staging, has been a promising method for building embedded domain-specific languages.
In such practices of multi-stage programming, a program usually is split into two stages: stage 1 (at compile time) executes the user code partially to generate an optimized intermediate program, and stage 2 (at run time) executes the intermediate program to get the final result.
In our design, programmers express the solvers as sparse matrices and their operations on stage 1, which generates matrix-free stage-2 code.

Lightweight Modular Staging (LMS) \cite{lms} is one of such frameworks and has been adopted in multiple DSL works \cite{delite,spiral,legobase}.
LMS separates the two stages through only types: \texttt{Rep[T]} refers to a \texttt{T}-typed value at the second stage.
Since that LMS is a Scala library, we implement our own to build our embedded DSL in Python.

\begin{figure}[t]
    \centering
    \begin{subfigure}[t]{0.35\textwidth}
        \centering
        \includegraphics[scale=0.5]{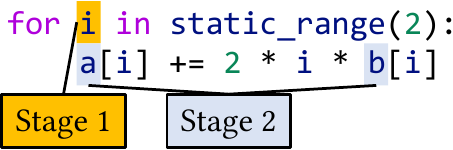}
        \subcaption{Code using type-based staging.}
        \label{fig:staging-bg:a}
    \end{subfigure}
    \hfill
    \begin{subfigure}[t]{0.6\textwidth}
        \centering
        \includegraphics[scale=0.5]{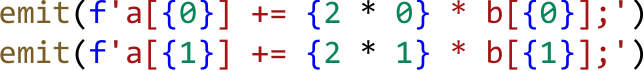}
        \subcaption{Logically equivalent code through manual string splicing.}
        \label{fig:staging-bg:b}
    \end{subfigure}
    \caption{A simple example for type-based multi-stage programming.}
    \label{fig:staging-bg}
    \vspace{-1em}
\end{figure}

In such an approach, expressions to be delayed to the next stage are used together with normal values.
The operators, including conventional ones (e.g. \texttt{\_\_mul\_\_(a, b)} for \texttt{a * b}) and indexed load/store (\texttt{\_\_getitem\_\_} and \texttt{\_\_setitem\_\_} for code like \texttt{a[...]}), are overloaded to make user code operating on the delayed expressions seamlessly.
When the first stage runs, the codes generated by the overloaded operators for the second stage are spliced together for later compiling.
The control flow structures are similarly overloaded to be able to generate control flow codes in the next stage.
In particular, our backend, built on top of \FT \cite{freetensor}, transforms the Python AST to enable virtualized control flow, the details of which will be introduced in \Cref{sec:staging:virt}.
A simple example is shown in \Cref{fig:staging-bg}.

\subsection{Polyhedral Analysis for Automatic Parallelization and Optimization}\label{sec:bg:par}

\begin{figure}[t]
    \begin{subfigure}[t]{0.48\textwidth}
        \centering
        \includegraphics[width=0.7\textwidth]{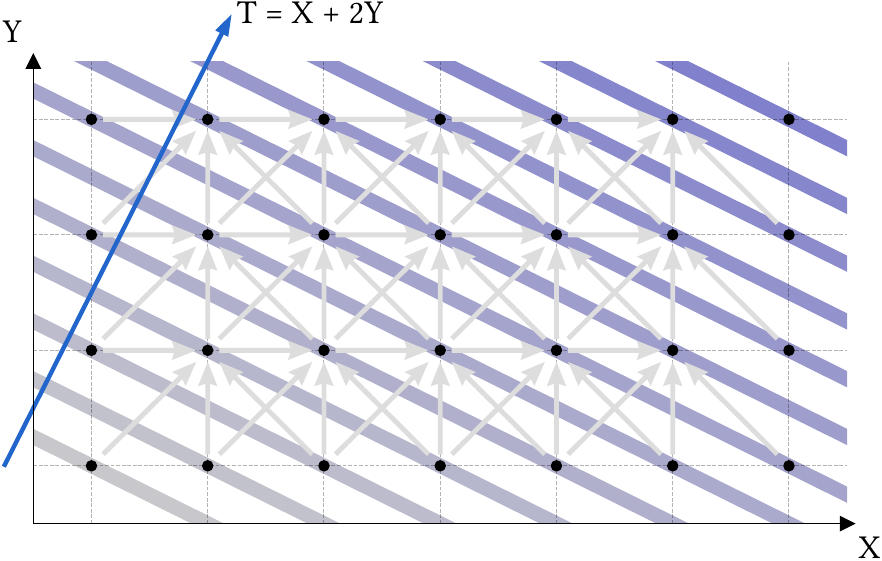}
        \subcaption{A typical skewing parallel implementation. Colored bands are executed in parallel, constructing a skewed ``time'' axis $T$. Poor locality.}
    \end{subfigure}
    \hfill
    \begin{subfigure}[t]{0.48\textwidth}
        \centering
        \includegraphics[width=0.7\textwidth]{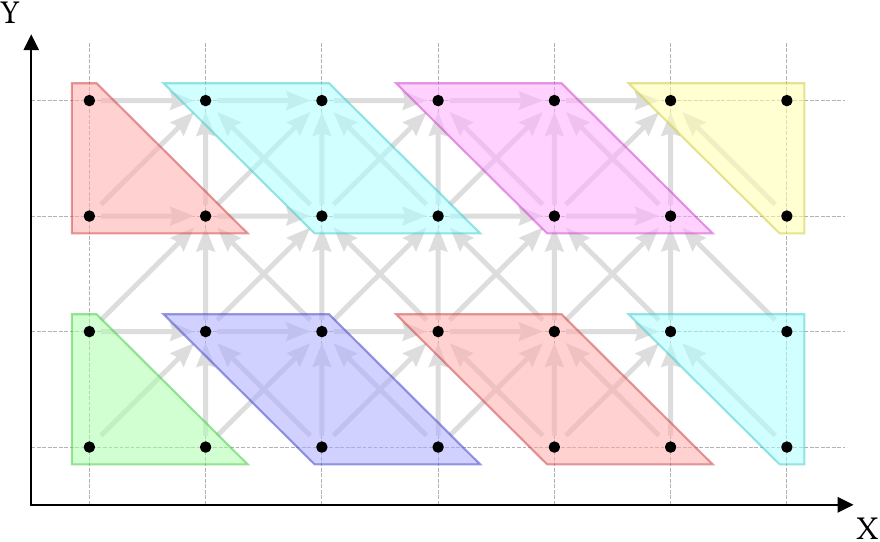}
        \subcaption{A typical skewed tiling parallel implementation. Tiles with the same color are executed in parallel, while each tile is executed sequentially by one thread. High startup overhead.}
    \end{subfigure}
    \vspace{-1em}
    \caption{Automatically-produced polyhedral schedules for 2D 9-point Gauss-Seidel.}
    \label{fig:seidel-existing}
\end{figure}

Through multi-stage programming, we produce matrix-free code containing the generalized stencils, in the form of quasi-affine tensor programs.
Such programs contain indexed for-loops and branches according to quasi-affine expressions, and access tensors, or multi-dimensional arrays, with quasi-affine indices as well.
There have been many studies on optimizing such affine programs since the invention of FORTRAN, which can be summarized as \textit{polyhedral analysis}.
Program analysis and transformations for multithreading, vectorization, and better cache management are summarized in \cite{DBLP:journals/cacm/PaduaW86} and \cite{DBLP:books/mk/AllenK2001}.
As such methods can only be applied on nested loops, they can hardly be used directly on domain-specific user code, but they point out a path towards automatically parallelizing our generated code.

The most famous ones among those compilers are PPCG \cite{ppcg} and PLUTO \cite{pluto}.
While capable of parallelizing computation in many different patterns, they need more specially-designed approaches to achieve efficient stencil computations, e.g. diamond tiling in PLUTO \cite{diamond}.
However, previous stencil optimizations mainly target Jacobi-style stencils, leaving Seidel-style stencils with vanilla skewing and/or tiling.
An example is shown in \Cref{fig:seidel-existing}, which contains two typical methods to parallelize a 2D 9-point Gauss-Seidel loop, produced by the existing automatic approaches.
The approach without tiling involves too many global synchronizations and has a poor memory locality.
The approach with skewed tiling instead incurs huge starting and ending overhead.
As such, they are not efficient at many of the implicit solvers our DSL wants to cover.
The development of new techniques resolving the Seidel-style stencils is thus essential for us.

\section{Overview of \SYS Language and Compiler}\label{sec:overview}

\begin{figure}[h]
    \centering
    \includegraphics[width=\textwidth]{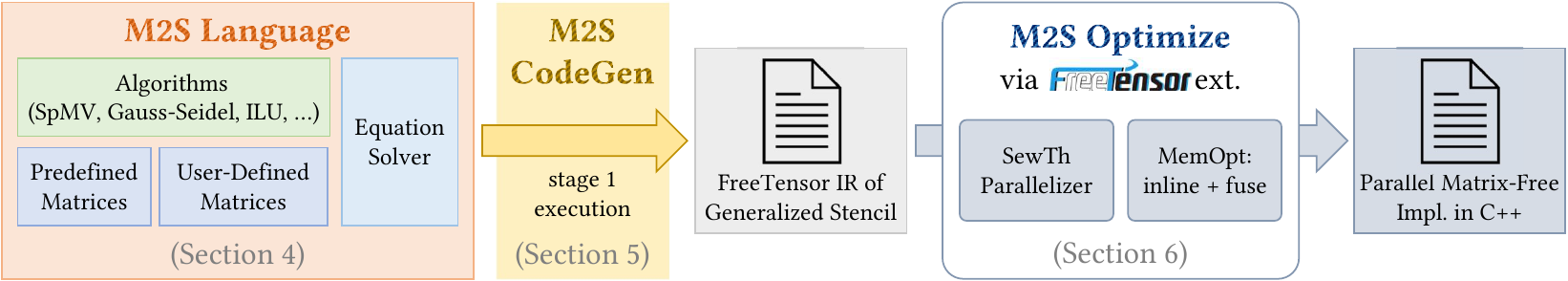}
    \caption{An overview of \SYS workflow.}
    \label{fig:lang-overview}
\end{figure}

Based on the observation mentioned above, we propose \SYS, our domain-specific language and its corresponding compiler for solving partial differential equations (PDEs) on structured grids.
An overview is shown in \Cref{fig:lang-overview}.
\SYS offers a concise and flexible programming interface for a wide range of PDE solvers.
Our structured sparse matrix abstraction enables the modular construction of sparse matrices through linear algebra operations on both predefined and user-defined simple matrices.
Furthermore, our DSL separates the algorithms from the matrices, supporting operations ranging from simple stencil operators to complicated Gauss-Seidel iterations and Incomplete LU decomposition.
Users write equation solver code that composes simple sparse matrices using arithmetic operations into complicated ones and invokes the algorithms against them.
The compile-time resolvable feature enables us to exploit sparsity patterns effectively.
The language design will be covered in \Cref{sec:ssr}.
As the first stage of the multi-stage programming frontend, we assemble matrix-free generalized stencil code in FreeTensor IR \cite{freetensor}, by backtracking and discussing over compile-time undecidable branches.
The staging techniques involved will be covered in \Cref{sec:staging}.

After the first stage, the backend optimization of \SYS, which we implemented as a \FT extension, is responsible for optimizing and parallelizing the code before proceeding to stage 2 (or runtime).
We propose a novel SewTh parallelizer (\textbf{Sew}ing the \textbf{Th}reads), which empowers the \SYS backend to automatically and efficiently parallelize Seidel-style stencils.
Additionally, we use a PLUTO+-like \cite{plutoplus} algorithm to fuse the loops, and perform aggressive inlining of elementwise functions, to optimize memory traffic further.
They will be discussed in \Cref{sec:backend}.

\section{\SYS Language Design}\label{sec:ssr}

In this section, we present the detailed design of the \SYS language.
We will start from the uppermost abstraction, which is on equation solvers, and then go down to the matrices abstraction and expression of linear algebra algorithms within \SYS.
Finally, we present the grammar of \SYS.

\subsection{Abstraction over the Equation}\label{sec:ssr:eq}

\begin{figure}[b]
    \begin{subfigure}[b]{0.6\textwidth}
        \centering
        \begin{align*}
            \left\{\mathbf{I} - \Delta\tau\left[
                \frac{\partial}{\partial x}\Lambda_x(\mathbf{u})
                + \frac{\partial^2}{\partial x^2}\rho N(\mathbf{u})
                - \epsilon h^4 \frac{\partial^4}{\partial x^4}
            \right]\right\}\ & \Delta \mathbf{u}_2 = \Delta \mathbf{u}_1 \\
            \left\{\mathbf{I} - \Delta\tau\left[
                \frac{\partial}{\partial y}\Lambda_y(\mathbf{u})
                + \frac{\partial^2}{\partial y^2}\rho Q(\mathbf{u})
                - \epsilon h^4 \frac{\partial^4}{\partial y^4}
            \right]\right\}\ & \Delta \mathbf{u}_4 = \Delta \mathbf{u}_3 \\
            \left\{\mathbf{I} - \Delta\tau\left[
                \frac{\partial}{\partial z}\Lambda_z(\mathbf{u})
                + \frac{\partial^2}{\partial z^2}\rho S(\mathbf{u})
                - \epsilon h^4 \frac{\partial^4}{\partial z^4}
            \right]\right\}\ & \Delta \mathbf{u}_6 = \Delta \mathbf{u}_5
        \end{align*}
        \subcaption{The equations to solve.}
        \label{fig:npb-sp-eq}
    \end{subfigure}
    \hfill
    \begin{subfigure}[b]{0.39\textwidth}
        \centering
        \includegraphics[width=\linewidth]{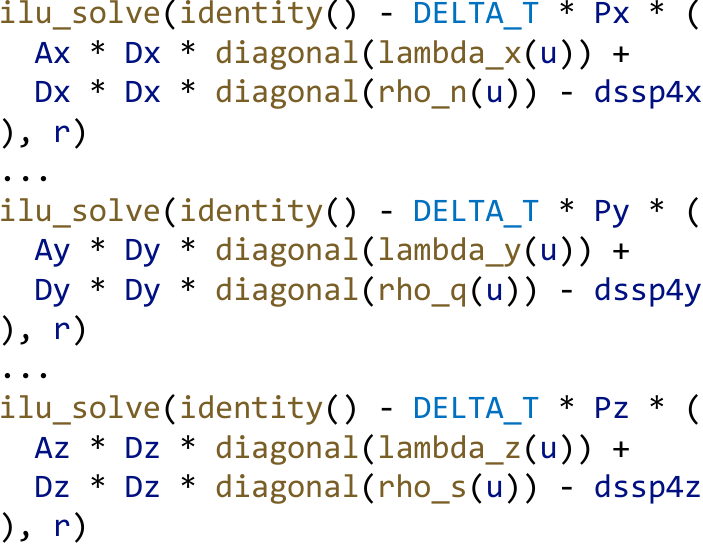}
        \caption{Solver implementation.}
        \label{fig:npb-sp-code}
    \end{subfigure}
    \caption{The NPB-SP specified problem and corresponded \SYS implementation.}
    \label{fig:npb-sp}
\end{figure}

In \SYS, the solving procedure of a PDE is expressed with sparse linear algebra, in which the sparse matrices involved are structured and modular.
As is mentioned in \Cref{sec:bg:sparse}, constructing the sparse matrix can be problematic for complicated real-world solvers.
Instead of requiring users to generate an entire matrix to solve with twisted problem generation code, we enable the assembly of such problem matrices through linear algebra expressions evaluated at compile-time.
We present how \SYS abstracts over the equation in \Cref{fig:npb-sp} with an end-to-end example from the Scalar Penta-diagonal (SP) pseudo-application in the NAS Parallel Benchmark (NAS) before we dig into the details.
Notice that instead of requiring the user to fill in the non-zeros of the matrices to solve one by one like in other matrix-based approaches, \SYS only requires users to implement individual modules, including custom structured sparse matrices like \texttt{dssp4} for $\epsilon h^4 (\partial^4/\partial *^4)$ and \texttt{lambda\_y}, \texttt{rho\_n}, etc. as elementwise functions, then write linear expressions to combine them, which are close to the discretized equations. 

One may notice that derivative operators like $\partial/\partial x$ are replaced by some variables like \texttt{Dx}.
These variables are predefined matrices that match the discretized differential and help the construction of matrices to solve.
We observe that the sparse matrices to solve can usually be constructed from several simple matrices, and the construction procedure is tightly related to the differential formula.
To explain this, we start with the first-order derivative.
After discretization, the $\partial/\partial x$ differential operator is turned into a sparse matrix $\mathbf{D}$, as is shown below.
For simplicity, we write the expanded form of a column vector $\mathbf{v}$ as a semicolon-separated list: $[v_0; ...; v_n]$.
\begin{align*}
    u
        \xrightarrow{discretize}& \left[u_0; ...; u_n\right] & \ \xlongequal{def}\  & \mathbf{u} \\
    \frac{\partial u}{\partial x}
        \xrightarrow{discretize}& \left[\frac{u_1 - u_0}{\Delta x}; ...; \frac{u_n - u_{n-1}}{\Delta x}\right] = \frac{1}{\Delta x}
        \begin{fbmatrix}
            -1 & 1 \\
            & \ddots & \ddots \\
            && -1 & 1 \\
        \end{fbmatrix}_{(n-1) \times n} \mathbf{u}
        & \ \xlongequal{def}\ & \mathbf{D}_{(n)} \mathbf{u}
\end{align*}

Based on that, we similarly have
$\frac{\partial^2 u}{\partial x^2}$ discretized as $\mathbf{D}_{(n-1)} \mathbf{D}_{(n)} \mathbf{u}$,
which can be simpler noted as $\mathbf{D}^2\mathbf{u}$ since the actual shape of each $\mathbf{D}$ can always be inferred from what vector it is multiplying to.
Beside that, we further define $\mathbf{P} = \left[\mathbf{0}; \mathbf{I}; \mathbf{0}\right]$, so that $\mathbf{P} \mathbf{u} = \left[0; \mathbf{u}; 0\right]$, which represents padding on the boundary.
Using this notation, we can easily perform the discretization in \Cref{eq:heat-eq-implicit} without dealing with scalars:
\begin{align}
    \nonumber &\frac{\mathbf{u}^{(T)} - \mathbf{u}^{(T-1)}}{\Delta t} = \left[0; \mathbf{D}^2 \mathbf{u}^{(T)}; 0\right] = \mathbf{P}\mathbf{D}^2 \mathbf{u}^{(T)} \\
    \Rightarrow\ \ &(\mathbf{I} - \Delta t \mathbf{P} \mathbf{D}^2) \mathbf{u}^{(T)} = \mathbf{u}^{(T-1)}. \label{eq:heat-eq-implicit-elementary}
\end{align}

It can be easily verified that \Cref{eq:heat-eq-implicit-elementary} is equivalent to \Cref{eq:heat-eq-implicit-mat}.
It naturally specifies how the matrix to solve is constructed, with a number of arithmetic operations on sparse matrices.
Furthermore, since that higher dimensional $\mathbf{u}$ will require flattening to a vector, $\mathbf{D}$ will be broadcasted into $\mathbf{D}_x$, $\mathbf{D}_y$, and $\mathbf{D}_z$, reflecting the higher-dimensional differencing operations.
Similarly, the neighborhood averaging $\mathbf{A}$, which has occurred in \Cref{fig:npb-sp-code} in broadcasted forms, is defined as
\begin{equation*}
    \mathbf{A} =
    \begin{fbmatrix}
        1/2 & 1/2 \\
        & 1/2 & 1/2 \\
        && \ddots & \ddots \\
        &&& 1/2 & 1/2 \\
    \end{fbmatrix}, \text{ so that }
    \mathbf{A} \mathbf{u} = \left[\frac{u_0 + u_1}{2}; ...; \frac{u_{n-1} + u_n}{2}\right].
\end{equation*}

However, computing the sparse matrices via these arithmetics at runtime can be costly, especially noticing that we need multiplications between sparse matrices.
Given the compile-time known structure of the structured sparse matrices, we have the opportunity to fully resolve the matrix construction at compile-time.
Besides, to generate efficient matrix-free compute kernels, the matrix-vector multiplication and sparse linear solvers need to cooperate with such structured matrices at compile time.
Given the multi-stage programming infrastructure, the first stage (running in Python) is responsible to deal with this.
The following subsections will introduce our design behind, and the staging procedure will be introduced in \Cref{sec:staging}.

\subsection{Sparse Matrices as Row Functions}\label{sec:ssr:rowfunc}

\begin{table}[t]
    \caption{Examples of sparse matrices in \SYS.}
    \label{tab:complicated-matrices}
    \begin{tblr}{colspec={|Q[c,m]|Q[c,h]|Q[l,h]|}, hlines, stretch=0}
        \SetCell[c=3]{l} \textbf{Predefined Matrices Examples} & & \\
        \begin{turn}{90}Diff\end{turn} &
        \includegraphics[align=c,width=0.12\textwidth]{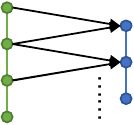} &
        \includegraphics[align=c,scale=0.5]{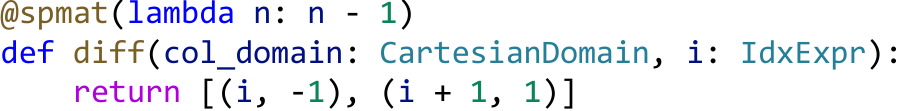} \\
        \begin{turn}{90}Padding\end{turn} &
        \includegraphics[align=c,width=0.12\textwidth]{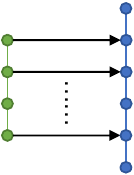} &
        \includegraphics[align=c,scale=0.5]{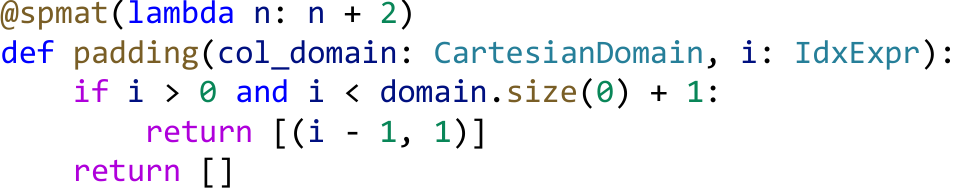} \\
        \SetCell[c=3]{l} \textbf{User-defined Matrices Examples} & & \\
        \begin{turn}{90}NPB 4th order dissipation\end{turn} &
        \includegraphics[align=c,width=0.12\textwidth]{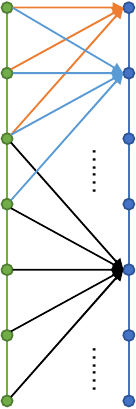} &
        \includegraphics[align=c,scale=0.5]{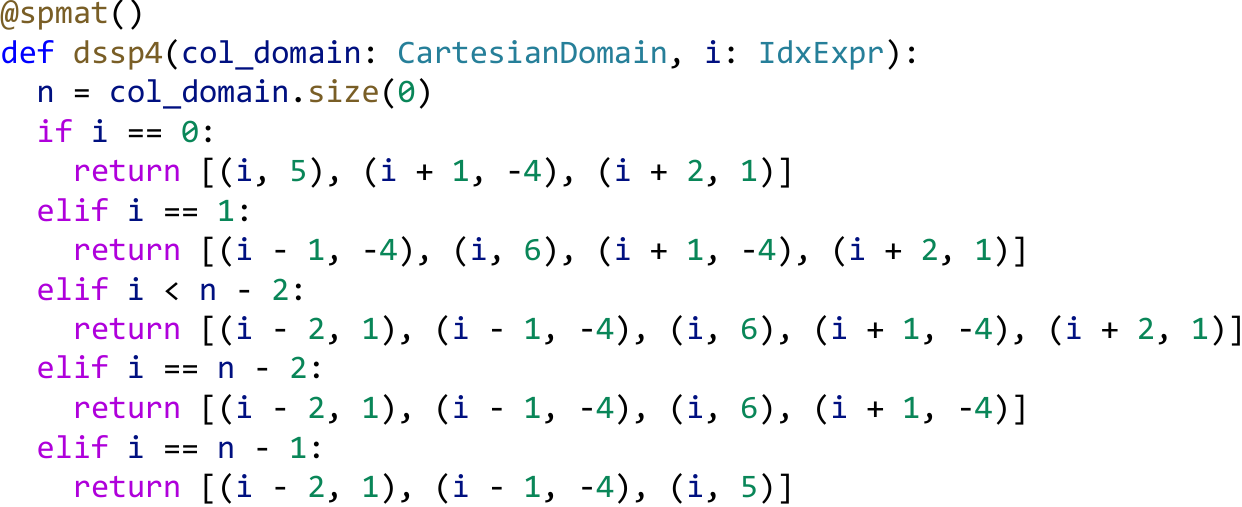} \\
        \begin{turn}{90}HPCG Prolong\end{turn} &
        \includegraphics[align=c,width=0.12\textwidth]{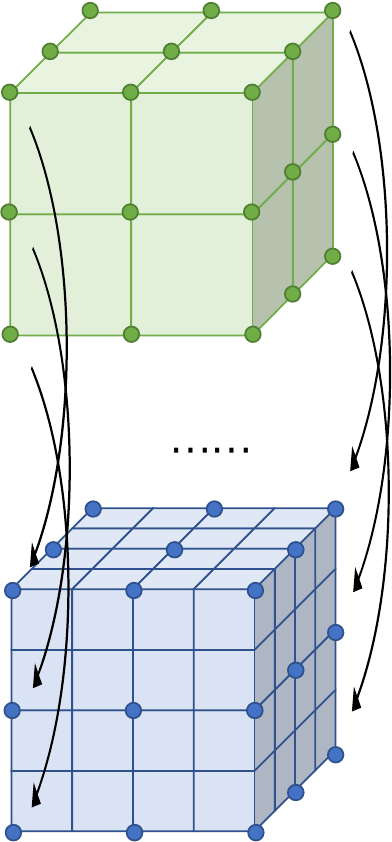} &
        \includegraphics[align=c,scale=0.5]{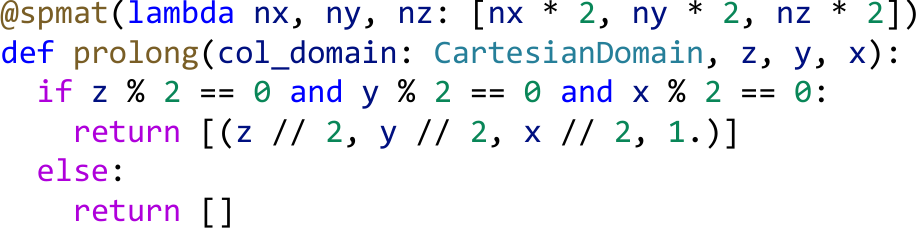} \\
    \end{tblr}
\end{table}

In \SYS, a sparse matrix is represented as a row function, which takes a row index and produces non-zeros in the specified row.
Each non-zero is a pair of a column index and the corresponding matrix value, in the same way as in the adjacency list.
The returned non-zero list is then consumed by sparse linear algebra procedures that define the computation, decoupled from the matrix.
Speaking in the multi-stage programming framework, the row index, non-zero column indices, and values are all staged expressions that will be resolved at runtime, while the non-zero list itself is a compile-time Python list.

Some examples are shown in \Cref{tab:complicated-matrices}.
A domain is assigned to both the row and column indices, representing their valid range starting from 0.
Thus, the row function also takes a parameter \texttt{col\_domain}, which contains the information on matrix size and will be described in detail in \Cref{sec:ssr:extend:md-domain}.
Decorated by \texttt{@spmat}, which is a Python decorator function we provide, the row function is wrapped into a \SYS matrix object of type \texttt{SpMat}.
The decorator also optionally takes a lambda expression to compute the matrix size relatively, the detail of which will be discussed in \Cref{sec:ssr:extend:adjustable}.
Besides, through simple if-else sequences, the special boundary in many structured sparse matrices can be well expressed, and so can periodical sparsity patterns that occur in, for example, prolongation in multi-grid solvers.

Our row function design imitates the adjacency list, corresponding the non-zero cells in the matrix to their rows.
This abstraction suits our applications well: an unknown variable in a discretized linear equation system like \Cref{eq:heat-eq-implicit} is related to a limited and compile-time known number of other unknowns, which are usually neighbors on the grid.
Thus in the matrix representing the linear systems to be solved, the non-zeros in a row are in the columns of the neighbors.
It makes the row function a natural design: the existing approach through general sparse linear algebra routines also requires users to generate sparse matrices with similar code, but for the more complex end-to-end matrices instead of our simple modular ones.

Yet, during implementing applications, we observe that expressing a conventional matrix is not sufficient to maximally ease user programming.
We extend the sparse matrix in \SYS in the following aspects and expose optimization opportunities.

\subsubsection{Multi-dimensional rows and columns}\label{sec:ssr:extend:md-domain}
Representing the structured problems as sparse matrices requires flattening the multi-dimensional unknowns.
It adds burdens on user programming to write the flattened index of the rows and columns.
Besides, such flattened access is not friendly to our backend optimization: to address the arbitrary dependence Seidel-style stencils may bring to us, the polyhedral analysis need complete information about the original multi-dimension structure to achieve good performance.

As such, we extend the sparse matrix to higher dimensions on the rows and columns.
Such a ``matrix'' with $m$-D row domain and $n$-D column domain is thus a $(m, n)$-tensor in the terminology of General Relativity; yet for convenience of understanding and to match the code, we will continue to note it as a ``matrix'', or \texttt{SpMat}.
A matrix object then has two properties, \texttt{row\_domain} and \texttt{col\_domain}
Each domain object has a tuple of integers for its grid size; in our current implementation, only regular Cartesian grids are considered, but the domain type can be extended easily.
An $m$-D point in the row domain corresponds to a row in the matrix consisting of non-zeros, each with a $n$-D point in the column domain and the value of the non-zero in the matrix.
Correspondingly, a previously flattened vector becomes a multi-dimensional \texttt{Vector} either, with respect to the multi-dimensional space that the physical quantities span over.
A \texttt{domain} is similarly attached as a property to each \texttt{Vector} object.

For example, in the prolongation matrix shown in \Cref{tab:complicated-matrices} we are already using the multi-dimensional row index passed into the row function and returning multi-dimensional column indices of non-zeros in the row.
If, for example, consumed by the SpMV procedure in \Cref{fig:spmv}, the returned column indices are later used to index the multi-dimensional \texttt{Vector} being multiplied against, thus performing the prolongation step mapping the coarser grid to a finer grid.

\subsubsection{Automatically deriving row domain from column}\label{sec:ssr:extend:adjustable}
A structured operation usually applies to many different grids, including grids of different sizes and dimensions.
We notice that a given column domain for a structured sparse matrix will decide its row domain.
In specific, a \texttt{SpMat} is logically right-multiplied to \texttt{Vector}s, thus the column domain is always decided first.
As such, we make the \texttt{row\_domain} of an \SYS matrix always following its \texttt{col\_domain}.
Walking into details, \texttt{SpMat} does not require specific row or column domains in user codes.
At each staging site that we decide its concrete shape, its \texttt{col\_domain} will be set through a Python property setter.
The setter calls a virtual function \texttt{set\_col\_domain} to set the column domain and compute the row domain.
The \texttt{@spmat} generated \texttt{SpMat} optionally receives a lambda expression as its argument to map its sizes of column domain to row, which is used to implement its \texttt{set\_col\_domain}.

\begin{figure}[b]
    \centering
    \hfill
    \begin{subfigure}[b]{0.27\textwidth}
        \centering
        \includegraphics[width=\textwidth]{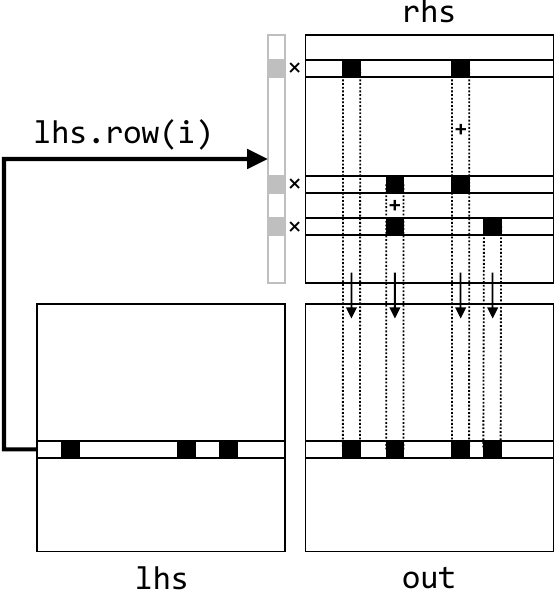}
        \subcaption{Row-based algorithm.}
    \label{fig:spgemm:a}
    \end{subfigure}
    \hfill
    \begin{subfigure}[b]{0.72\textwidth}
        \centering
        \vspace{-1em}
        \includegraphics[width=\textwidth]{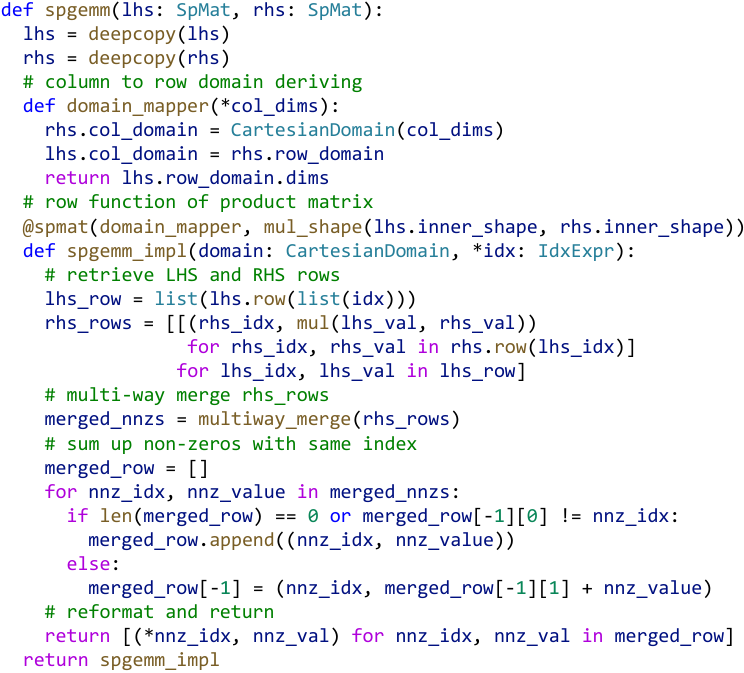}
        \subcaption{Implementation in \SYS.}
    \label{fig:spgemm:b}
    \end{subfigure}
    \hfill
    \vspace{-2em}
    \caption{Algorithm and implementation of SpGEMM.}
    \label{fig:spgemm}
\end{figure}

\subsubsection{Inner value as a tensor}\label{sec:ssr:extend:tensor-inner}
Multiple (but a small fixed number of) quantities are often attached to the same grid in real-world applications.
In previous matrix-based approaches, these quantities are flattened along with the grid dimensions; after retaining the grid dimensions, we then need to allow the \texttt{SpMat} and \texttt{Vector} to have tensors, instead of plain scalars, as its inner values.
They thus have a fixed \texttt{inner\_shape}, which describes the shape of their inner tensor values, as an extra parameter in constructing a \SYS sparse matrix.

\subsection{Linear Algebra on \SYS}

Through the programming interfaces of \SYS, linear algebra routines can be easily implemented.
Wrapped in a domain loop, such routines can retrieve lists of non-zeros in each row of \SYS matrices, and perform computation according to their algorithm.
Those sparse linear algebra routines are able and expected to be implemented in their simplest form, i.e. directly operating on the adjacency list, without the need for manual optimizations.

\begin{wrapfigure}{r}{0.35\linewidth}
    \includegraphics[width=\linewidth]{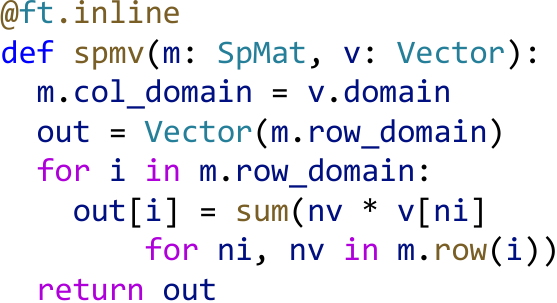}
    \caption{Implementation of SpMV in \SYS.}
    \vspace{-1em}
    \label{fig:spmv}
\end{wrapfigure}

\subsubsection{Sparse matrix-vector multiplication (SpMV)}
The simplest example is SpMV in \Cref{fig:spmv}, multiplying a sparse matrix and a vector.
It first adjusts the domains of the matrix and retrieves the expected domain of the product vector accordingly, the mechanism of which has been covered in \Cref{sec:ssr:extend:adjustable}.

Looping over the domain, it reads corresponding rows through the \texttt{row} method of the matrix object, which is a simple wrap of the above row function.
It then multiplies the value from the matrix and vector for each non-zero and sums them up to get the final result, following the math of matrix-vector multiplication.

\subsubsection{Sparse matrix multiplication}
With both operands expressed in \SYS, it is possible to give their product still in \SYS and being resolved all at compile stage through a row-wise \texttt{spgemm} as is shown in \Cref{fig:spgemm:a}.

Then, as shown in \Cref{fig:spgemm:b}, the row function of the resulting matrix first walks through the row of the specified index on the left-hand side (LHS), retrieving the non-zeros.
It then uses the non-zeros' column indices to index the rows of the right-hand side (RHS) matrix.
All the RHS rows are merged in order, multiplied with the LHS values accordingly, and summed up to one non-zero sequence, producing the result row.
All this procedure happens in stage 1 and is thus expanded to plain matrix-free codes before stage 2.

Summing up and subtracting between \SYS matrices is even simpler: they only require merging two rows from the two matrices, similar to combining the multiple rows from RHS in \texttt{spgemm}.
We also implement them for \SYS but do not include details here.

\begin{wrapfigure}{r}{0.46\textwidth}
    \includegraphics[width=\linewidth]{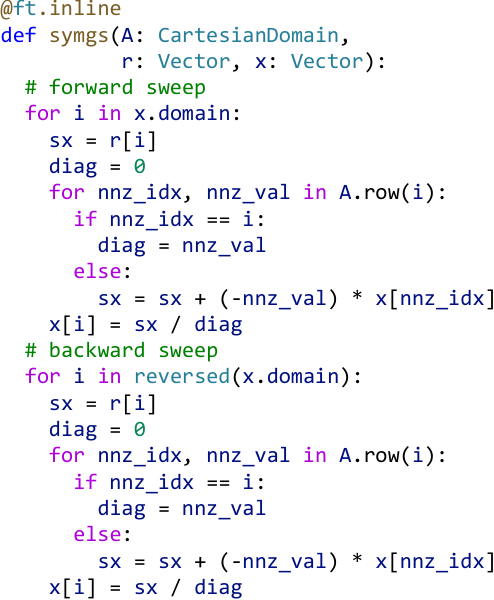}
    \caption{Example code for Symmetric Gauss-Seidel implemented against \SYS.}
    \label{fig:symgs}
\end{wrapfigure}

\subsubsection{Enabling spatial dependence}
These linear algebra routines can also contain spatial dependence carried by the domain loops, according to their different algorithms, enabling the implementation of more complicated routines required by PDE solvers.
An example is in \Cref{fig:symgs}, in which the loops simultaneously read and write on the same \texttt{Vector} object and introduces matrix-defined dependence.
The staging procedure will then produce their matrix-free form constructed of regular nested loops, without any indirect access typical in traditional sparse linear algebra, thanks to the staged domain loop indices (will be covered in \Cref{sec:staging}).
Our analysis and optimization mentioned in \Cref{sec:backend} will resolve the loop-carried dependence and exploit chances to parallelize.

To support our evaluated applications, we have implemented Incomplete Lower-Upper, Symmetric Successive Over-Relaxation, and Symmetric Gauss-Seidel, each with tens of lines of code.
Users with science backgrounds will be able to implement their own routines in \SYS in usually tens of lines, much easier than in traditional sparse algebra libraries.

\subsection{Grammar of \SYS}\label{sec:lang:grammar}

After describing our language design, we thereby demonstrate the grammar of the \SYS language.
As an embedded domain-specific language built upon type-based staging techniques, \SYS involves various host language (Python) functions during the staging process, noted as $\text{InType} \rightsquigarrow \text{OutType}$ and may have side effects.
Many designs mentioned above work as Python components, and are thus invisible in the grammar, including the row function, structured sparse matrix objects, and linear algebra routines.
Without hurting the core idea, it also omits details about ``inner value as a tensor'' from \Cref{sec:ssr:extend:tensor-inner}.

\begin{align*}
S_O \Coloneqq &   && \mathbf{new\_vector} (\text{domain}:\text{List}[\text{Int}], \text{vector}:\text{Name}) & \text{(creating vector on domain)} \\
        & | && \mathbf{new\_scalar} (\text{scalar}:\text{Name}) & \text{(creating global scalar)} \\
        & | && \mathbf{store\_scalar} (\text{scalar}:\text{Name}, \text{expr}:E_O) & \text{(store to global scalar)}\\
        & | && \mathbf{for} (\text{domain}:\text{List}[\text{Int}], \text{body}:\text{List}[I] \rightsquigarrow S_I) & \text{(runtime loop, dictionary order)} \\
        & | && \mathbf{ite}_O (\text{cond}:B_O, \text{then}:S_O, \text{else}:\text{Optional}[S_O]) & \text{(runtime branch)} \\
        & | && S_O; S_O & \text{(statements sequence)} \\
S_I \Coloneqq &   && \mathbf{store} (\text{vector}:\text{Name}, \text{idx}:\text{List}[I], \text{expr}:E_I) & \text{(store an element to vector)}\\
        & | && \mathbf{reduce\_add} (\text{scalar}:\text{Name}, \text{expr}:E_I) & \text{(reducing to global scalar)} \\
        & | && \mathbf{ite}_I (\text{cond}:B_I, \text{then}:\rightsquigarrow S_I, \text{else}:\text{Optional}[\rightsquigarrow S_I]) & \text{(runtime branch)} \\
        & | && S_I; S_I & \text{(statements sequence)} \\
E_O \Coloneqq &   && \text{Float} & \text{(lift from compile-time)} \\
        & | && E_O + E_O \mid E_O - E_O \mid E_O \times E_O \mid E_O \ /\ E_O \mid ... & \text{(floating point arithmetics)} \\
        & | && \mathbf{load\_scalar} (\text{scalar}:\text{Name}) & \text{(load value of scalar)} \\
B_O \Coloneqq &   && E_O > E_O \mid E_O \geq E_O \mid E_O < E_O \mid E_O \leq E_O & \text{(floating point comparison)} \\
        & | && B_O \land B_O \mid B_O \lor B_O \mid \lnot B_O & \text{(boolean operations)} \\
E_I \Coloneqq &   && E_O & \text{(lift from global)}\\
        & | && E_I + E_I \mid E_I - E_I \mid E_I \times E_I \mid E_I / E_I \mid ... & \text{(floating point arithmetics)} \\
        & | && \mathbf{load} (\text{vector}:\text{Name}, \text{idx}:\text{List}[I]) & \text{(load element from vector)} \\
        & | && I & \text{lift from an integer index)} \\
I   \Coloneqq &   && \text{Int} \mid I + I \mid I - I \mid I * I \mid I \div I \mid I \text{ mod } I & \text{(integer literal and arithmetics)} \\
        & | && \text{Name} & \text{(index variable of domain loop)} \\
B_I \Coloneqq &   && I > I \mid I \geq I \mid I < I \mid I \leq I \mid I = I \mid I \neq I & \text{(integer comparison)} \\
        & | && B_I \land B_I \mid B_I \lor B_I \mid \lnot B_I & \text{(boolean operations)}
\end{align*}

Note that we support separate classes of statements and expressions for outside and inside domain loops.
Domain loops cannot nest with each other;
loading and storing the vectors only happens inside domain loops;
and the branches inside domain loops, $\mathbf{ite}_I$ must be conditioned according to integer arithmetics of integer indices $I$ ultimately introduced by domain loops.
Global scalars are useful to aggregate values globally, e.g., dot products.
The outer if-else-then $\mathbf{ite}_O$ is useful for iterative solver algorithms to control the workflow, e.g. check convergence.
The aforementioned sparse linear algebra routines, invoked with our structured sparse matrices, are all taken place outside the domain loops and compose $S_O$ nodes.

Then a sparse matrix will be a host language (Python) object of type:
\begin{align*}
\mathbf{spmat} (& \text{domain\_col2row}:\text{List}[\text{Int}] \rightsquigarrow \text{List}[\text{Int}], \\
& \text{row}:(\text{col\_domain}:\text{List}[\text{Int}], \text{row\_idx}:\text{List}[I]) \rightsquigarrow \\
& \hspace{2.43em} (\text{eff}:S_I, \text{nnzs}:\text{List}[(\text{col\_idx}:I, \text{value}:E)]))
\end{align*}

Which contains a function \texttt{domain\_col2row}, mapping column domain to row domain, and a function \texttt{row}, for retrieving a row of non-zeros in the matrix symbolically.

\section{Matrix-free Code Generation via Staging}\label{sec:staging}

\subsection{Customizable Control Flow Virtualization}\label{sec:staging:virt}

Similar to LMS, our staging infrastructure works through control flow virtualization.
With a decorator, it automatically transforms a function to its virtualized version: each for-loop and if-then-else branch is transformed into a corresponding virtual call, allowing the code generator to emit stage-2 code wrapping the loop or branch body.
But unlike LMS, which directs all control flow to a central overload object defining the stage-2 language, we choose to make the virtual call point to the iterable or predicate expression itself to gain more customizability.

We demonstrate examples of the code transformation in \Cref{tab:into-staging}.
Walking into the details, if the iterable or predicate is of staged type (\texttt{StagedIterable} and \texttt{StagedPredicate} accordingly), the staging handler method of the staged object (\texttt{foreach} and \texttt{if\_stmt} accordingly) is called to stage future codes.
Otherwise, it automatically falls back to a stage-1 control flow.
The logical operations, which are not overloadable per the design of Python, are virtualized similarly for \texttt{StagedPredicate}.
This staging infrastructure has been merged into FreeTensor as the basis of its frontend.

\begin{table}[b]
    \centering
    \caption{Examples of code transformation for control flow virtualization. Standalone temporary function definitions are used in real implementation instead of lambda expressions to enable multi-line bodies.}
    \label{tab:into-staging}
    \begin{tabular}{p{0.25\textwidth}p{0.7\textwidth}}
        \toprule
        \textbf{User code} & \textbf{After control flow virtualization} \\
        \midrule
        \midrule
        \begin{tabminted}{python}
for i in it:
  foo(i)
\end{tabminted}
        &
        \begin{tabminted}{python}
if isinstance(it, StagedIterable):
  it.foreach('i', lambda i: foo(i))
else:
  for i in it:  foo(i)
\end{tabminted}
        \\
        \hline
        \begin{tabminted}{python}
pred1 and pred2
\end{tabminted}
        &
        \begin{tabminted}{python}
(pred1.logical_and(pred2) if isinstance(pred1, StagedPredicate)
                          else (pred1 and pred2))
\end{tabminted}
        \\
        \hline
        \begin{tabminted}{python}
pred1 or pred2
\end{tabminted}
        &
        \begin{tabminted}{python}
(pred1.logical_or(pred2)  if isinstance(pred1, StagedPredicate)
                          else (pred1 or pred2))
\end{tabminted}
        \\
        \hline
        \begin{tabminted}{python}
not pred
\end{tabminted}
        &
        \begin{tabminted}{python}
(pred.logical_not()       if isinstance(pred, StagedPredicate)
                          else (not pred))
\end{tabminted}
        \\
        \hline
        \begin{tabminted}{python}
if pred:
  foo()
else:
  bar()
\end{tabminted}
        &
        \begin{tabminted}{python}
if isinstance(pred, StagedPredicate):
  pred.if_stmt(lambda: foo(), lambda: bar())
else:
  if pred:  foo()
  else:     bar()
\end{tabminted}
        \\
        \bottomrule
    \end{tabular}
\end{table}

It is natural to see that the domain loops iterating through all points in a domain have to be staged as loops in stage 2.
Otherwise, the staged code size will bloat out as the problem grows, which is undesired since we will deal with large science problems.
As such, we implement \texttt{Staged}-\texttt{Iterable} for the domain objects and emit the nested for-loops according to the domain dimensions in the staging handler.
It also allows us to do more work in the staging handler behind the scenes to support the backtracking and discussing which we will introduce next.

\subsection{Resolve undecidable branches through backtracking and discussing}\label{sec:staging:backtrack}

\begin{figure}[t]
    \centering
    \includegraphics[width=\linewidth]{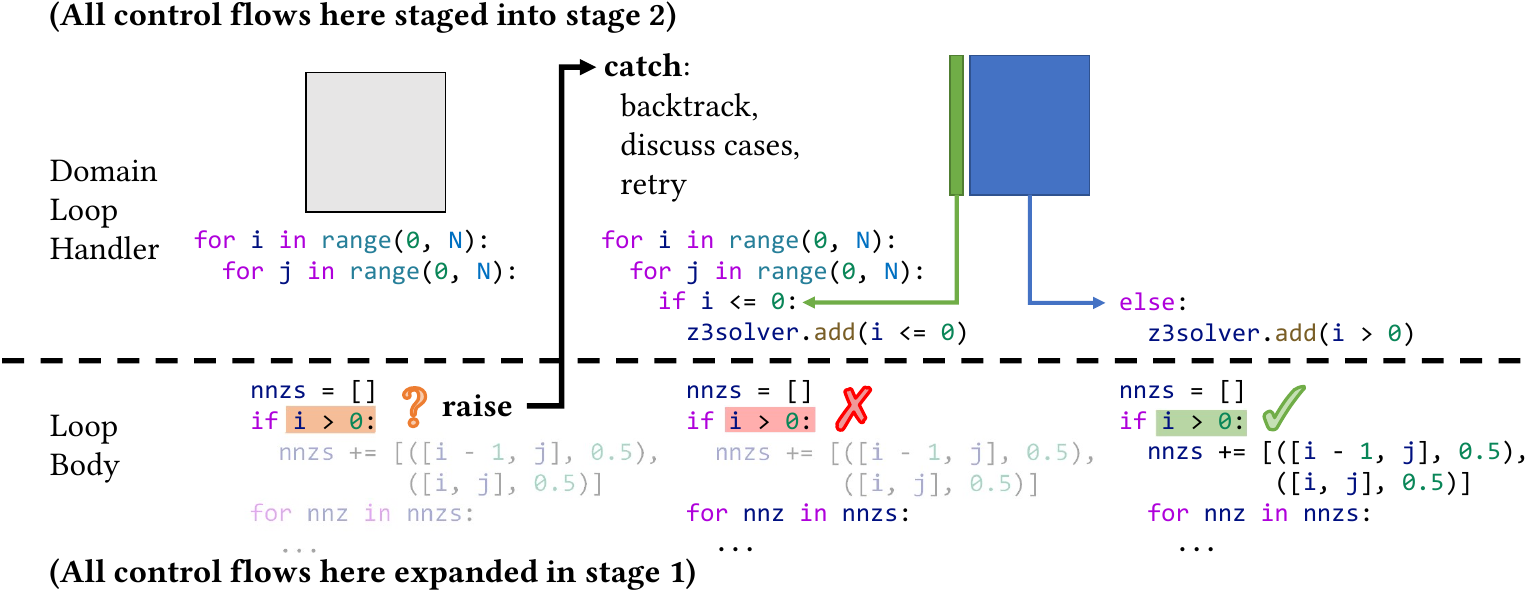}
    \caption{The procedure of backtracking and discussing for resolving undecidable branches.}
    \label{fig:backtrack}
\end{figure}

As we have shown in \Cref{sec:ssr:rowfunc}, we want to enable taking branches with regard to row indices and manipulating stage-1 lists accordingly.
Besides, in matrix multiplication, we need similar functionality to merge non-zero sequences according to the comparison results of column indices; more user programs also need such branches.
They together raise a vital requirement on branching in stage 1 according to expressions related to the domain loop indices.
Formally speaking, the $\mathbf{ite}_I$ node shown in \Cref{sec:lang:grammar} accepts a runtime boolean expression $B_I$ but coordinates with compile-time then/else bodies as Python functions, which is the time-reversing.
However, notice that possible combinations of these branches are available at stage 1, we can split the original loop domain into pieces, making control flows deterministic in a single piece.

\algblockdefx[Try]{Try}{EndTry}%
    {\textbf{try}}%
    {}
\algcblockdefx[Catch]{Try}{Catch}{EndTry}%
    {\textbf{catch}}%
    {}
\algtext*{EndIf}  
\algtext*{EndFor} 
\algtext*{EndWhile} 
\algtext*{EndFunction} 
\begin{algorithm}[t]
\caption{Pseudo-code for the backtrack and discuss algorithm.}
\label{alg:backtrack}
\begin{algorithmic}[1]
    \Function{$S_O$.for}{$\textit{domain}: \text{List}[\text{Int}], \textit{body}: \text{List}[I]$}
        \Let{\textbf{global} \textit{solver}}{Z3.Solver()} \Comment{create global solver}
        \Let{\textit{indices}}{$nil$} \Comment{create list of loop indices}
        \For{$n \gets \textit{domain}$}
            \Let{$i$}{create new Name as $I$}
            \State $\textit{solver}.\text{add}(0 \leq i < n)$ \Comment{append constraint for dimension range}
            \Let{\textit{indices}}{$\textit{indices} :: i$} \Comment{append loop index of this dimension}
        \EndFor
        \Function{RecursingBacktrack}{}
            \Try
                \State \Return{$\textit{body}(\textit{indices})$}
            \Catch\ Undeterminant(\textit{cond})
                \State \textit{solver}.push$(\textit{cond})$ \Comment{discuss with \textit{cond} holds}
                \Let{\textit{true\_case}}{\textsc{RecursingBacktrack}$()$}
                \State \textit{solver}.pop$()$
                \State \textit{solver}.push$(\lnot\textit{cond})$ \Comment{discuss with \textit{cond} not holds}
                \Let{\textit{false\_case}}{\textsc{RecursingBacktrack}$()$}
                \State \textit{solver}.pop$()$
                \State \Return{\texttt{FT.ite}$(\textit{cond}, \textit{true\_case}, \textit{false\_case})$} \Comment{assemble the wrapping if-else-then}
            \EndTry
        \EndFunction
        \Let{\textit{code}}{\textsc{RecursingBacktrack}$()$}
        \For{$i \gets \text{len}(domain)-1 .. 0$} \Comment{assemble the domain for loops}
            \Let{\textit{code}}{\texttt{FT.for}$(\textit{indices}[i], \texttt{FT.slice}(\textit{domain}[i]), \textit{code})$}
        \EndFor
        \State \Return{\textit{code}}
    \EndFunction
    \State
    \Function{$S_I$.$\text{ite}_I$}{$\textit{cond}:B_I, \textit{then}:\rightsquigarrow S_I, \textit{else}:\text{Optional}[\rightsquigarrow S_I]$}
        \State \textbf{global} \textit{solver} \Comment{retrieve the global solver}
        \If{$\textit{solver}.\text{prove}(\textit{cond})$} \Comment{use \textit{then} when always true}
            \State \Return{$\textit{then}()$}
        \ElsIf{$\textit{solver}.\text{prove}(\lnot\textit{cond})$} \Comment{use \textit{else} if exists when always false}
            \State \Return{\textbf{if} \textit{else} \textbf{then} $\textit{else}()$ \textbf{else} \texttt{FT.nop}}
        \Else \Comment{undeterministic, raise exception to discuss the cases}
            \State \textbf{raise} Undeterminant$(\textit{cond})$
        \EndIf
    \EndFunction
\end{algorithmic}
\end{algorithm}

To achieve this, we backtrack on every undecidable branch and discuss them to be true or false.
As is shown in \Cref{fig:backtrack}, we backtrack through raising an exception when an undeterministic predicate is met in the $\mathbf{ite}_I$ if-else-then branch.
Such branches must occur inside some domain loop, the stage 1 handler of which will catch the exception and retry staging the loop body twice, each with the predicate assumed to be true or false.
The time-reversing branches thus become regular runtime branches.
We implement the boolean expressions $B_I$ as a sub-class of \texttt{StagedPredicate} and implement the checking and raising logic in its handler to achieve this.

It can be seen that as more branches are met, the number of cases easily grows exponentially with the number of branches.
We observe that the predicates in \SYS codes for structured solving turn out to be primarily duplicated; it is reasonable since the domain dimension is limited to no more than 3 in typical usages, and the branches are mostly related to boundaries or grid coarsening so the number of independent predicates on each dimension will be limited to only a handful.
As such, we further utilize the Z3 \cite{z3} SMT solver to deduplicate the cases.
Every time a predicate is taken to generate the loop body for its case, we append the assumption into the Z3 solver.
Once a new predicate is met, we first check its truth value through Z3; if deterministic, we will not perform backtracking and instead directly take the branch.
Applying such a scheme, we are able to generate clean case-discussing code for matrix-free algorithms.

Summarizing above, the backtrack and discuss algorithm is shown in \Cref{alg:backtrack}.

\section{Backend Optimization}\label{sec:backend}

In this section, we introduce our backend techniques to parallelize and optimize the generated matrix-free code.
The most significant part is SewTh, a novel scheduling algorithm to efficiently parallelize arbitrary Seidel-style stencil operators.
Unlike typical polyhedral schedulers, which linearly transform nested loops, optionally tile them, and selectively parallelize, SewTh additionally schedules fine-grained point-to-point synchronizations at certain iterations instead of performing global barrier synchronization.
We also implement elementwise function inlining and loop fusion with linear transformation based on the Pluto+ algorithm \cite{plutoplus}.

\subsection{Sewing the Threads: A New Approach to Parallelize Seidel-style Stencils}

\begin{figure}[t]
        \vspace{1em}
    \begin{subfigure}[t]{0.65\textwidth}
        \centering
        \includegraphics[width=\textwidth]{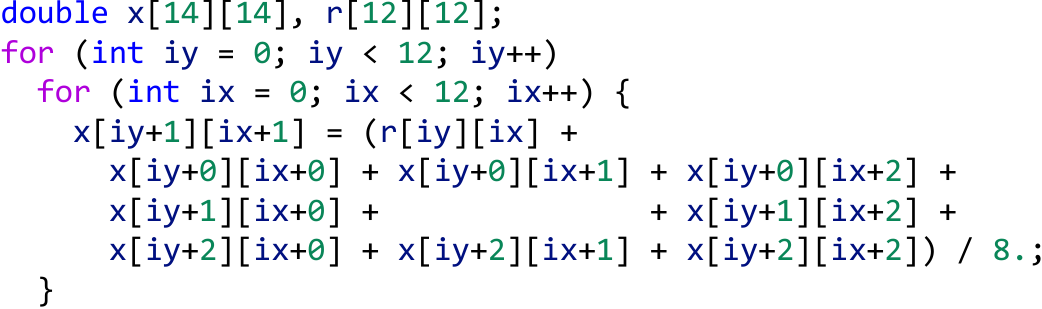}
        \subcaption{Input code of a 2D 9-point Gauss-Seidel.}
        \label{fig:sewth-input}
    \end{subfigure}
    \begin{subfigure}[t]{0.65\textwidth}
        \vspace{1em}
        \centering
        \includegraphics[width=\textwidth]{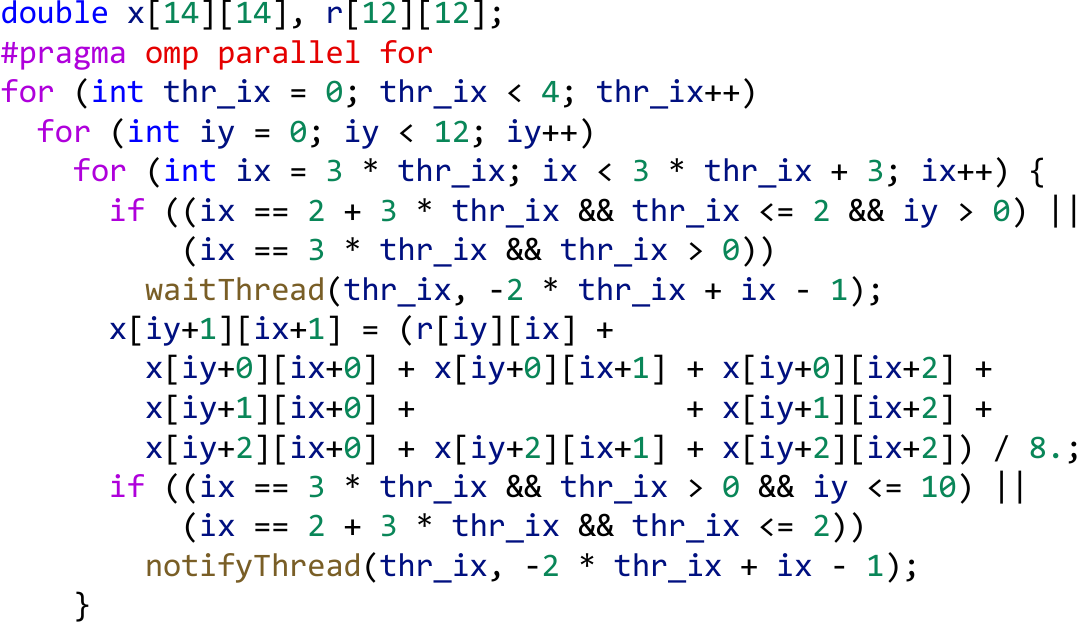}
        \subcaption{Generated code by SewTh.}
        \label{fig:sewth-codegen}
    \end{subfigure}
    \begin{subfigure}[t]{0.98\textwidth}
        \vspace{1em}
        \centering
        \includegraphics[width=0.7\textwidth]{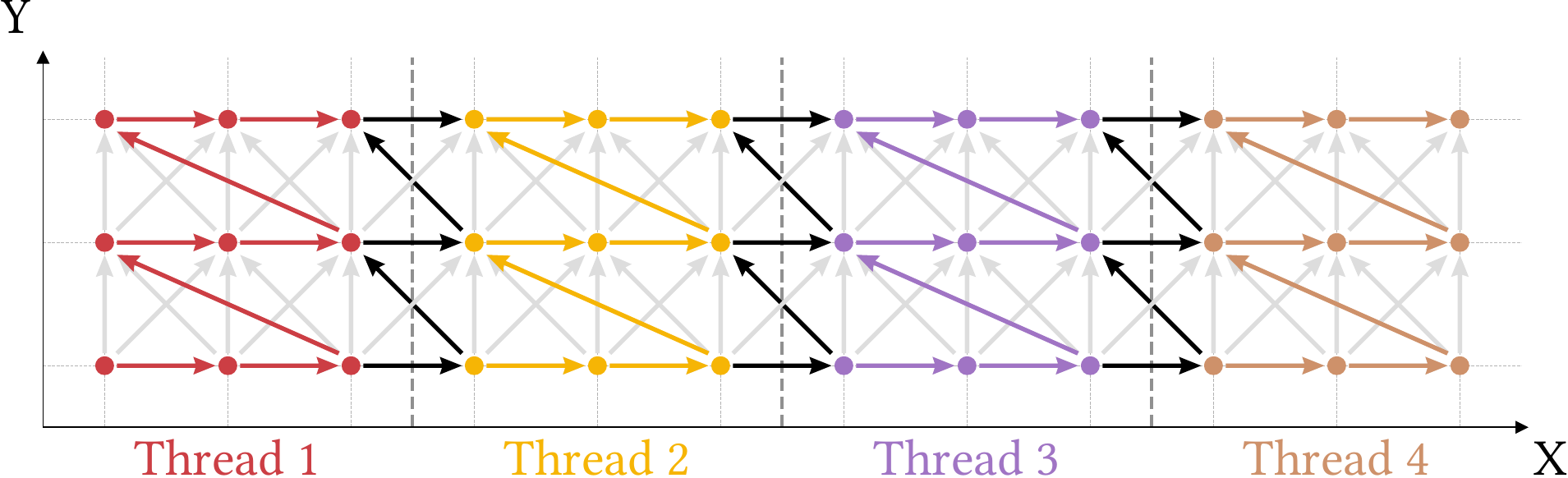}
        \subcaption{The enforced partitioning schedule.
        The sequences of different colored arrows are in sequential execution order within threads.
        The arrows crossing thread boundaries are the \textbf{crossing dependence}.
        Dark arrows are the \textbf{compacted crossing dependence}, which are the only physically synchronized dependence.}
        \label{fig:enforced-tiling}
    \end{subfigure}
    \caption{An example of SewTh that parallelizes a 2D 9-point Gauss-Seidel operator. Equivalent C++ code is presented instead of \FT IR for better readability.}
\end{figure}

As is mentioned in \Cref{sec:bg:par}, automatic parallelization of Seidel-style Stencils has not been explored extensively by existing research.
To achieve both good memory locality and low overhead starting, our core idea is to assign part of the loop to each thread and perform fine-grained synchronization to satisfy the dependence across threads.
Unlike the existing polyhedral code generation approaches synchronizing with global barriers, we generate point-to-point synchronizations based on single-producer, single-consumer (SPSC) lock-free queues.
They are thus on-demand and introduce significantly lower overhead.
With the large blocks assigned to each thread, it is rare to perform the synchronization.
Such fine-grained synchronization sews the threads together, guaranteeing the correctness of produced code.

We cover the details of the work decomposition and what to synchronize, how we compute necessary synchronization from analyzed dependence and generate code for them, and how we automatically prove the parallelism of our plan.
We will use the 2D 9-point Gauss-Seidel presented in \Cref{fig:sewth-input} as the example throughout this subsection.
And we will continue to abbreviate $\left[u_1, u_2, \ldots, u_n\right]$ as bold $\mathbf{u}$.

\subsubsection{Enforced partitioning to threads}\label{sec:backend:sewth:tile}

To exploit parallelism from the multidimensional loops with spatial dependence, we first assign certain partitioning to the loop space.
We partition as few dimensions as possible in all inner loops without any skewing, while leaving the outermost loop and possibly some inner loops not partitioned.
Assume the dimension sizes of the outer domain loops are described by $\mathbf{N}$.
Organizing the threads into a mesh, their partition sizes are $\mathbf{N}^t$, of length 1 less than $\mathbf{N}$, selected so that components of $\mathbf{N}^t$ are close to each other.
Then we enforce the function mapping an iteration $\mathbf{x}$ to a thread to be $\mathbf{T}(\mathbf{x}) = \left[{x_{1}}\div{N^t_0}, \ldots, {x_{i+1}}\div{N^t_i}, \ldots\right]$, in which $\mathbf{x} \in L = \{\mathbf{x} \mid \forall i, 0 \leq x_i < N_i\}$ (the looping range.)
Within each thread, it still takes the lexicographical execution order, so that the intra-thread dependence naturally holds.
An example is shown in \Cref{fig:enforced-tiling}.


It is then only the dependence crossing thread boundaries, or \textit{crossing dependence}, remained to be considered.
The crossing dependence can be computed from the full dependence set, which is extracted from the program with the help of polyhedral dependence analysis and Presburger arithmetics.
To start with, the full dependence set computed by FreeTensor dependence analysis is noted as $D = \{(\mathbf{a} \in L, \mathbf{b} \in L) \mid \mathbf{b} \; \text{depends on} \; \mathbf{a}\}$.
For all $(\mathbf{a}, \mathbf{b}) \in D$, $\mathbf{b}$ must be computed later than $\mathbf{a}$ in any valid schedule.
The formal definition of crossing dependence is shown in \Cref{def:cross-dep}, which is also the Presburger arithmetics to automatically compute it.

\begin{definition}
\label{def:cross-dep}
The crossing dependence, in the form of a Presburger map from thread pair to dependencies, is computed as below:
\[
D_{\times} (\mathbf{T}_a, \mathbf{T}_b) = \{(\mathbf{a}, \mathbf{b}) \in D \mid \mathbf{T}(\mathbf{a}) = \mathbf{T}_a \land \mathbf{T}(\mathbf{b}) = \mathbf{T}_b \land \mathbf{T}_a \neq \mathbf{T}_b\}.
\]
\end{definition}

Synchronizing the crossing dependence guarantees two depending iterations belonging to different threads to be executed in order.
It is obvious that such synchronization will not cause deadlock: both the intra-thread execution order and inter-thread synchronization are following the lexicographical order of the multi-dimensional loop indices (also the execution order for sequential input program), making the combined synchronization graph acyclic.

\subsubsection{Compact dependence and generate code}
We have addressed above what dependence should be covered by inter-thread synchronization.
Yet, directly generating synchronizing code according to the crossing dependence has multiple downsides.
Firstly, simply putting one point-to-point synchronization for each crossing dependence works fine, but introduces redundant synchronization.
Besides, the possibly differently ordered sources and targets between a thread pair require the generated code to accurately synchronize against specific loop iterations instead of simply synchronizing against the thread.
Taking the case in \Cref{fig:enforced-tiling}, we attempt to avoid the explicit synchronization on the tinted crossing dependence.

\begin{wrapfigure}{r}{0.25\linewidth}
    \vspace{-1em}
    \centering
    \includegraphics[width=\linewidth]{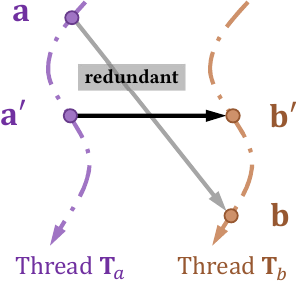}
    \caption{Compact crossing dependence.}
    \vspace{-2em}
    \label{fig:sewth-compact}
\end{wrapfigure}

We propose to compact the crossing dependence as follows.
A crossing dependence is considered to be redundant if there is another crossing dependence between the same thread pair with a later or same source and an earlier or same target.
Non-redundant crossing dependence is retained during compaction, i.e. compacted crossing dependence.
A visual illustration is shown in \Cref{fig:sewth-compact}:
since the sequential order guarantees $\mathbf{a}'$ happens later than $\mathbf{a}$, $\mathbf{b}$ later than $\mathbf{b}'$, and another dependence $\mathbf{a}' \rightarrow \mathbf{b}'$ exists, $\mathbf{a} \rightarrow \mathbf{b}$ is marked redundant and not synchronized explicitly.
The formal definition (which is also the algorithm) is listed in \Cref{def:compact-dep}.

\begin{definition}
\label{def:compact-dep}
The redundant crossing dependence is computed as:
\begin{align*}
D_{\times}^- (\mathbf{T}_a, \mathbf{T}_b) = \{(\mathbf{a}, \mathbf{b}) \in D_{\times}(\mathbf{T}_a, \mathbf{T}_b) \mid \exists (\mathbf{a}', \mathbf{b}') \in D_{\times}(\mathbf{T}_a, \mathbf{T}_b),\, & \left(\mathbf{a} < \mathbf{a}' \land \mathbf{b}' \leq \mathbf{b}\right) \lor \\
 & \left(\mathbf{a} = \mathbf{a'} \land \mathbf{b'} < \mathbf{b}\right)\},
\end{align*}

in which the comparisons are carried out in dictionary order.
Then the compacted crossing dependencies are:
\[
D_{\times}^+ (\mathbf{T}_a, \mathbf{T}_b) = D_{\times} (\mathbf{T}_a, \mathbf{T}_b) - D_{\times}^- (\mathbf{T}_a, \mathbf{T}_b).
\]
\end{definition}

Then we conclude the following properties we want:

\begin{theorem}[Completeness]\label{theorem:completeness}
With the sequential execution order of each thread and the compacted crossing dependence satisfied by point-to-point synchronization, all crossing dependence are guaranteed to hold. Formally speaking, we have
\[
\forall (\mathbf{a}, \mathbf{b}) \in D_{\times} (\mathbf{T}_a, \mathbf{T}_b),\, \exists (\mathbf{a}', \mathbf{b}') \in D_{\times}^+ (\mathbf{T}_a, \mathbf{T}_b),\, \mathbf{a} \leq \mathbf{a}' \land \mathbf{b}' \leq \mathbf{b}.
\]
\end{theorem}

\begin{proof}
We will prove this theorem by contradiction.

Suppose for some $(\mathbf{a}_0, \mathbf{b}_0) \in D_{\times} (\mathbf{T}_a, \mathbf{T}_b)$,
$ \forall (\mathbf{a}', \mathbf{b}') \in D_{\times}^+ (\mathbf{T}_a, \mathbf{T}_b), \mathbf{a}_0 > \mathbf{a}' \lor \mathbf{b}' > \mathbf{b}_0. $

If $(\mathbf{a}_0, \mathbf{b}_0) \in D_{\times}^+ (\mathbf{T}_a, \mathbf{T}_b)$, then taking $(\mathbf{a}', \mathbf{b}') = (\mathbf{a}_0, \mathbf{b}_0)$ contradicts the assumption.

Otherwise, according to \Cref{def:compact-dep}, we have $(\mathbf{a}_0, \mathbf{b}_0) \in D_{\times}^- (\mathbf{T}_a, \mathbf{T}_b)$.
Thus, $$\exists (\mathbf{a}_1, \mathbf{b}_1) \in D_{\times} (\mathbf{T}_a, \mathbf{T}_b), (\mathbf{a}_0, \mathbf{b}_0) \neq (\mathbf{a}_1, \mathbf{b}_1) \land \mathbf{a}_0 \leq \mathbf{a}_1 \land \mathbf{b}_1 \leq \mathbf{b}_0.$$

We can then repeat the process and get a sequence of different dependence $(\mathbf{a}_2, \mathbf{b}_2), (\mathbf{a}_3, \mathbf{b}_3), \dots$, until reaching some $(\mathbf{a}_n, \mathbf{b}_n) \in D_{\times}^+ (\mathbf{T}_a, \mathbf{T}_b)$.
The dependence set is finite, so we can always drain the $D_{\times}^- (\mathbf{T}_a, \mathbf{T}_b)$ and arrive at $D_{\times}^+ (\mathbf{T}_a, \mathbf{T}_b)$, resulting in the contradiction above.
\end{proof}

\begin{theorem}[Can Blindly Synchronize]\label{theorem:blindness}
For every thread pair, the sources and targets of the compacted crossing dependence are in the same order, so blindly synchronizing against the thread will be equivalent to synchronizing against the accurate iteration. Formally speaking, we have
\[
\forall (\mathbf{a}_1, \mathbf{b}_1) \text{ and } (\mathbf{a}_2, \mathbf{b}_2) \in D_{\times}^+ (\mathbf{T}_a, \mathbf{T}_b),\, \mathbf{a}_1 < \mathbf{a}_2 \Leftrightarrow \mathbf{b}_1 < \mathbf{b}_2.
\]
\end{theorem}

\begin{proof}
We will prove this theorem by contradiction.
Suppose there exists $(\mathbf{a}_1, \mathbf{b}_1), (\mathbf{a}_2, \mathbf{b}_2)$ in $D_{\times}^+ (\mathbf{T}_a, \mathbf{T}_b)$, that $\mathbf{a}_1 < \mathbf{a}_2 \Leftrightarrow \mathbf{b}_1 < \mathbf{b}_2$ does not hold.
Then according to \Cref{def:compact-dep} we have
\begin{align*}
(\mathbf{a}_2 \leq \mathbf{a}_1 \land \mathbf{b}_1 < \mathbf{b}_2) \;\lor\; & (\mathbf{a}_1 < \mathbf{a}_2 \land \mathbf{b}_2 \leq \mathbf{b}_1) \\
\Rightarrow (\mathbf{a}_2, \mathbf{b}_2) \in D_{\times}^- (\mathbf{T}_a, \mathbf{T}_b) \;\lor\; & (\mathbf{a}_1, \mathbf{b}_1) \in D_{\times}^- (\mathbf{T}_a, \mathbf{T}_b).
\end{align*}
Since $D_{\times}^+ (\mathbf{T}_a, \mathbf{T}_b) \cap D_{\times}^- (\mathbf{T}_a, \mathbf{T}_b) = \emptyset$, this contradicts the assumption.
\end{proof}

Then the compacted crossing dependence set between each thread pair can be computed as a quasi-affine map with the help of ILP libraries, ISL \cite{isl} in our case, allowing us to generate code against it.
A ``wait'' block is injected before the loop body so that it waits for each source thread that the current loop iteration depends on, and similarly, a ``notify'' block but inversed after the loop body.
According to \Cref{theorem:completeness}, the compaction ensures that only synchronizing the compacted dependence will be sufficient.
Also, according to \Cref{theorem:blindness} the synchronization can ignore which loop iteration the other end is and blindly synchronize with the specific thread.
We are thus possible to synchronize simply through efficient atomic counters, one for each thread pair.
A code generation example is shown in \Cref{fig:sewth-codegen}.

\begin{wrapfigure}{r}{0.3\linewidth}
    \vspace{-1.7em}
    \centering
    \includegraphics[width=\linewidth]{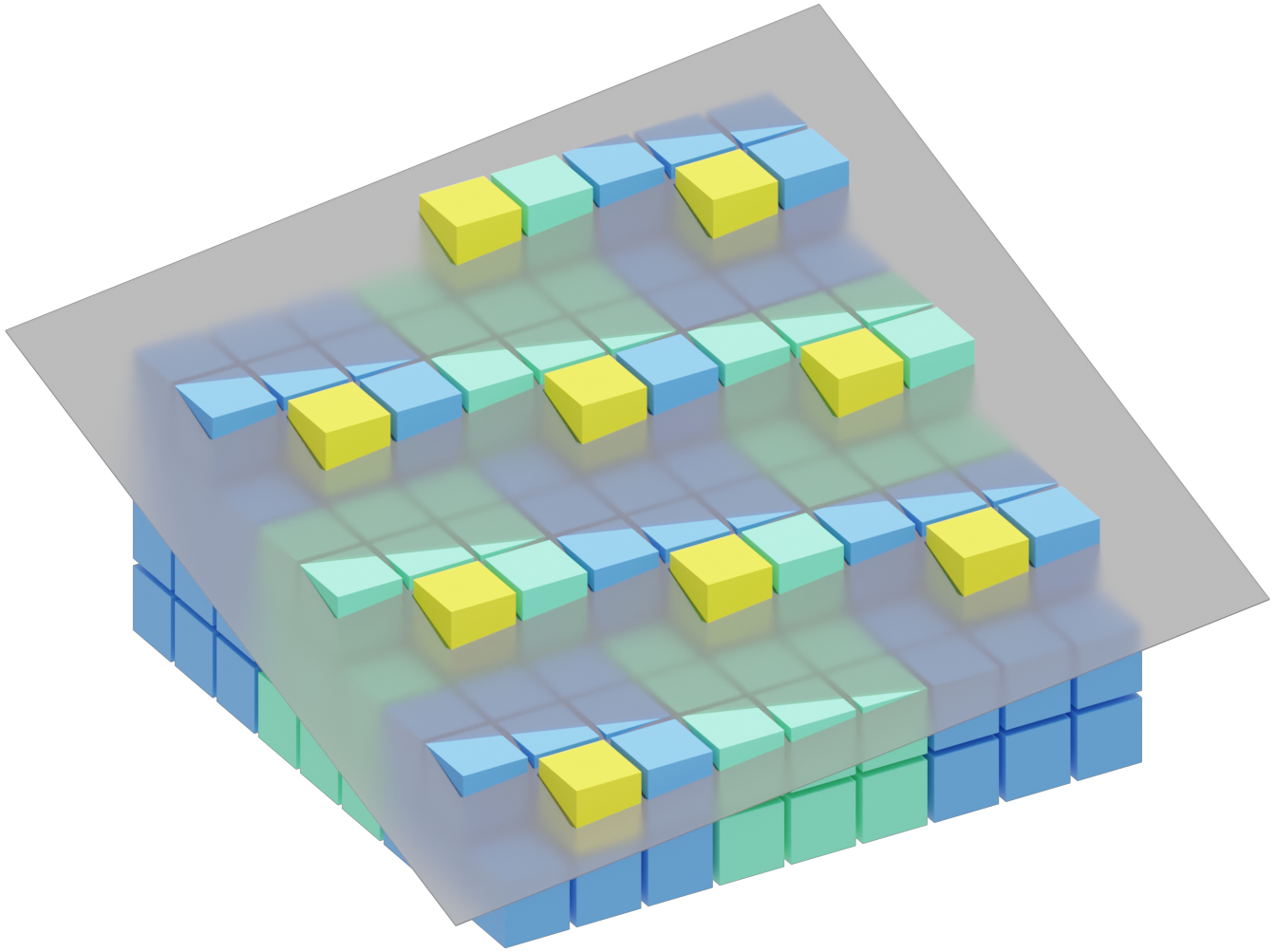}
    \caption{A plane from \cref{eq:hyperplane}.}
    \label{fig:hyperplane}
\end{wrapfigure}

\subsubsection{Modeling the parallelism}
The above work decomposition and synchronization together do not necessarily imply sufficient parallelism.
It is possible that most threads are busy waiting and only some of the threads are working all the time under inappropriate parameters.
We here propose a model for the parallelism lower bound of the generated code.

\begin{figure}[b]
    \centering
    \begin{subfigure}{0.48\textwidth}
        \centering
        \includegraphics[width=\linewidth]{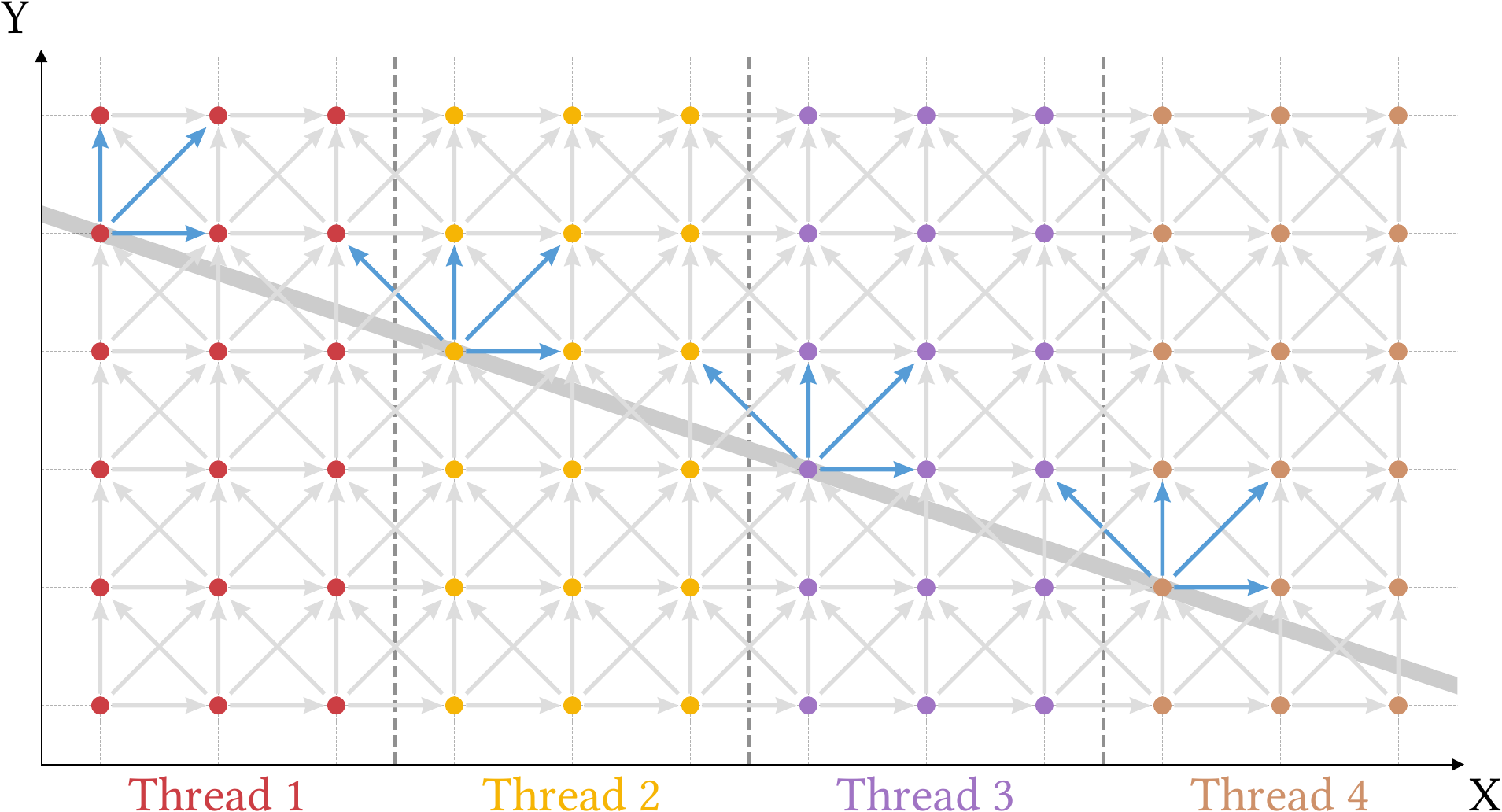}
        \subcaption{Good parallelism passing the thread partitioning hyperplanes check.}
        \label{fig:2d-hyperplane:legal}
    \end{subfigure}
    \hfill
    \begin{subfigure}{0.48\textwidth}
        \centering
        \includegraphics[width=\linewidth]{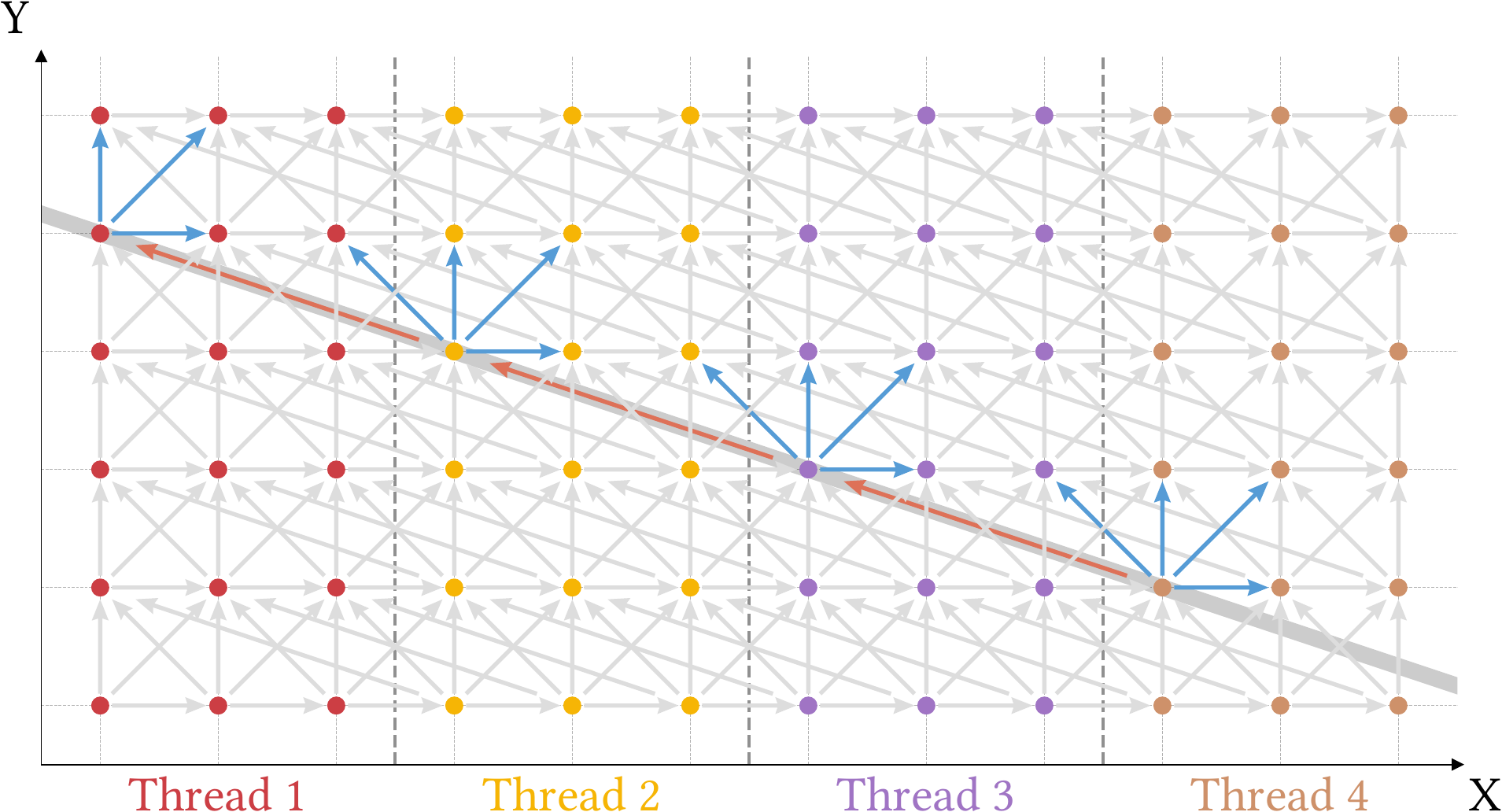}
        \subcaption{Degraded parallelism detected by the thread partitioning hyperplanes.}
        \label{fig:2d-hyperplane:illegal}
    \end{subfigure}
    \caption{Two examples for checking parallelism with the thread partitioning hyperplanes.}
    \label{fig:2d-hyperplane}
\end{figure}

Existing methods have been utilizing \textit{hyperplanes} to determine the time axis that enables parallelism.
Recall the parallelizable bands in \Cref{fig:seidel-existing}, these are an example of a valid setting of parallelizable hyperplanes: since the nested loop has 2 dimensions, the hyperplanes become 1D lines.
Inspired by those methods, we design a set of thread partitioning hyperplanes corresponding to the enforced partitioning proposed above.
Say we have the loop dimensions $\mathbf{x}=\left[x_0, x_1, ..., x_n\right]$, $0 \leq x_i < N_i$, and their inner partition sizes $\left[N^t_1, N^t_2, ..., N^t_n\right]$.
Then we define the thread partitioning hyperplanes as
\begin{equation}\label{eq:hyperplane}
\phi_t(\mathbf{x}) = \sum_{i=0}^{n} \prod_{j=i+1}^{n} N^t_j x_i = c.
\end{equation}
Such a partitioning hyperplane ensures contiguous execution on each thread, following the thread-local lexicographic execution order as specified in \Cref{sec:backend:sewth:tile}.
A visual example is shown in \Cref{fig:hyperplane}.
Each cube cell represents a loop iteration and chessboard-colored ranges are thread partitions.
The yellow cells are the loop iterations executed in parallel if the plane satisfies all loop dependence.

If such a hyperplane setting satisfies all dependence, then all loop iterations in the same hyperplane and assigned to different threads will be parallelizable.
Formally speaking, we check whether for any dependence pair $(\mathbf{a}, \mathbf{b}) \in D$, $\phi_t(\mathbf{a}) < \phi_t(\mathbf{b})$, as a sufficient condition for the parallelism.
If the check passed, the hyperplanes will reveal a possible parallel execution plan, and serve as a lower bound for the parallelism of our generated parallel implementation, which is already fully parallelized except at the short pipeline start and exit.

Take the two dimensional examples in \Cref{fig:2d-hyperplane}, in which $N^t_1 = 3$, so that $\phi_t(\left[ y, x \right]) = 3y+x$.
If the thread partitioning hyperplanes successfully satisfy all dependence as is in \Cref{fig:2d-hyperplane:legal}, our plan is guaranteed to provide near-optimal parallelism (fully parallel except at the beginning and the end).
Instead in \Cref{fig:2d-hyperplane:illegal}, the thread partitioning hyperplanes do not satisfy all dependence: for the brown dependence, $\phi_t(\left[ y + 1, x - 3 \right]) = \phi_t(\left[ y, x \right])$;
in such circumstances, our plan will yield degraded parallelism.

There is a great chance that the loop dependence in a seidel-style stencil operator satisfies this condition.
Given a large grid and usually tens of threads, the thread partitions will be much larger than the stencil pattern, the size of which is determined by the order of partial differential equation and usually a small constant.
The hyperplanes can then cover all iterations in the near neighborhood and earlier than the current iteration, satisfying all possible dependence.
For example, in \Cref{fig:2d-hyperplane:illegal}, the hyperplanes fail only after the long dependence is added.
The dependence is too long such that it spreads over an entire thread partition.
With larger loop spaces and stencils still in such sizes, it will be impossible.
Actually, in all evaluation cases requiring Seidel-style stencils in \Cref{sec:eval} including micro-kernels and end-to-end solvers, the thread partitioning hyperplanes satisfy all dependence in every kernel.

\subsection{Memory optimizations}

In addition to basically parallelizing everything, including Jacobi and Seidel-style stencils, we introduce memory optimizations as below.
While simple to work with, these optimizations are critical to many real-world applications, since the stencils involved in PDE solving are usually memory-bound, making reducing memory traffic important.

\subsubsection{Aggressive inlining of pointwise operations}
In \Cref{sec:ssr:eq} we have mentioned elementwise functions, which allow the user to easily build pointwise operations.
Using such components will create intermediate variables that can be easily inlined and eliminated.
We perform such inlining whenever possible so that the memory traffic is minimized.
The inlining is automatically accomplished by recognizing all elementwise domain loops and calling the \FT schedule \texttt{inline}.

\subsubsection{Loop fusion with affine transformation}
Loop fusion plays a critical role in memory intensive compute kernels, which is usually the case for the stencils in PDE solving.
While previous DSLs can perform loop fusion only based on operator-level dependence information, our added support for customizable spatial loop dependence prohibits such simple approaches.
Besides, the modern multigrid methods involve finer and coarser grids, and the loops on them can only be fused with strides.
As such, we implement automatic loop fusion also based on polyhedral dependence analysis, which iteratively selects one axis from each of the two adjacent loop nests, performs affine transformation computed similar to the PLUTO+ \cite{plutoplus} algorithm, and fuses the two loops; the procedure is repeated until no level of loops can be fused in the loop nests.
At the program level, the loop nests are iteratively fused, scanned in the forward order, and only fuse loops that do not sacrifice parallelism after fusion.
The implementation is accomplished by ourselves instead of invoking PLUTO+ directly.

\par

We do not perform tiling additionally since SewTh already exposes a tiled execution order.
Other than that, we also automatically vectorize sequential loops with best efforts through the corresponding \FT schedule \texttt{vectorize}; \FT determines if a loop is vectorizable also through polyhedral analysis.

\section{Evaluation}\label{sec:eval}

In this section, we evaluate our domain-specific language design and performance through end-to-end benchmarking and backend optimization assessment.
We set up our experiments on a dual-socket server, each socket being one Intel Xeon Gold 6126 CPU, with 12 physical cores or 24 logical cores per CPU, counting 24 physical or 48 logical together.
The server has 565 GB of DDR4-2666 main memory.
The server runs Ubuntu 18.04.4 LTS with Linux kernel 4.15.0.
We use GCC 12.2.0 to compile the programs, using \texttt{g++} for C++ and \texttt{gfortran} for FORTRAN, including our generated codes and baseline manual implementations.
All parallelism is through OpenMP, including reference codes and our implementation.

\subsection{End-to-end Benchmarking}

We evaluate \SYS against official manual implementations of all the three pseudo-appli-cations in the NAS Parallel Benchmark (NPB) \cite{npb}, BT, SP, and LU, which are three solvers of a Computational Fluid Dynamics (CFD) problem, and a matrix-free variant of High Performance Conjugate Gradient (HPCG) \cite{hpcg}, which is a widely adopted performance benchmark solving Poisson equation with a multigrid solver.

We use NPB 3.4.2 as the baseline for the NPB evaluation and a hand-written matrix-free HPCG from \cite{newsw-hpcg} for the HPCG evaluation.
Existing stencil DSLs are not capable of expressing these solvers due to the presence of Seidel-style stencils in them, thus we do not include such baselines.
We neither compare with the reference HPCG implementation, given it is matrix-based according to the specification and such comparison will be unfair\footnote{We test a 48-process reference HPCG run on a total of $768^3$ sized grid, which is $192\times192\times256$ local grid on each process, because the reference HPCG code does not support multithreaded SymGS.
It is worth noting that although it achieves full parallelization through domain decomposition, it is at the cost of worse convergence.
Even so, it takes $355.8$ seconds per run (50 iterations), which is significantly slower than $64.2$ seconds in \SYS, being $5.5\times$ longer.}.
For NPB, we run the Class B, C, D, and E problems (grid size $102^3$, $162^3$, $408^3$, and $1020^3$) on all three pseudo-applications, the sizes of which are specified in the NPB Specification.
For HPCG, we run on grids of size $768^3$, $1024^3$, $1280^3$, and $1536^3$.
We show the computed grid points per second, which is the grid size divided by time consumption.

\begin{figure}[h]
    \centering
    \includegraphics[width=\textwidth]{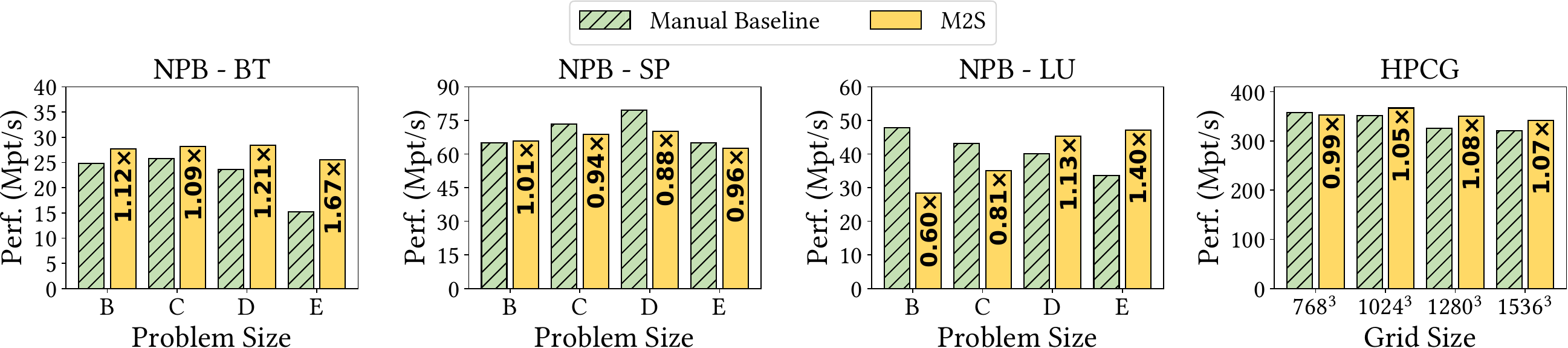}
    \caption{Performance results on NAS Parallel Benchmarks (NPB-BT, NPB-SP, NPB-LU), and HPCG. \textbf{Higher is better.}}
    \label{fig:eval}
\end{figure}

The evaluation result is shown in \Cref{fig:eval}.
In the three pseudo-applications of NPB, we achieve an average (geometric mean) of $1.25\times$, $0.94\times$, and $0.93\times$ performance, compared to the manual matrix-free implementation, respectively.
In HPCG, we achieve an average (geometric mean) of $1.04\times$ performance.
The performance is competitive at all scales.
We also observe that larger grids yield relatively better performance in our implementation, possibly due to that the Python-native interface overhead is less significant at a larger scale.

\subsubsection{Comparing the Code Length}

We evaluate the human effort required to program matrix-free solvers manually or using our \SYS by comparing the lines of code for different implementations in \Cref{tab:loc}.
\SYS implementation takes 1044 lines of Python code for the majority of work and 329 lines of C++ code as a \FT plugin implementing some auxiliary passes.
Based on that, we implement all three NPB pseudo-applications in a total of 750 lines of code, with 485 lines among them to be shared across applications.
It takes only $6.3\%$ of the total code length compared to the total of three NAS-provided manually matrix-free implementations.
Notably, even considering the lines of code for our compiler, we still involve significantly less code.
In addition, the matrix-free HPCG implementation through \SYS takes $16.4\%$ of the code length in hand-written matrix-free implementation.
\SYS significantly releases the burden of programming efficient matrix-free solvers.

\begin{table}[ht]
    \begin{minipage}{\textwidth}
        \centering
        \caption{Lines of code in evaluated cases under different implementations.}
        \label{tab:loc}
        \begin{tabularx}{0.95\textwidth}{XcX}
        &
        \begin{tabular}{c|ccc|c}
            \diagbox{Impl.}{App.} & NPB-BT & NPB-SP & NPB-LU & HPCG \\
            \hline
            Manual & 4288 & 3079 & 3068 & 914 \\
            \SYS & 408\textsuperscript{\textdagger}\footnotetext{\textsuperscript{\textdagger} Common codes for all NPB pseudo-applications; mostly elementwise functions with long math formulas.} + 24 & 408\textsuperscript{\textdagger} + 188\textsuperscript{\textdaggerdbl}\footnotetext{\textsuperscript{\textdaggerdbl}Longer than others due to its unique extra math formulas, mentioned in \Cref{sec:ssr:eq}.} & 408\textsuperscript{\textdagger} + 41 & 150 \\
        \end{tabular}
        &
        \end{tabularx}
    \end{minipage}
\end{table}

\subsubsection{Evaluating Compile time}

To demonstrate the compilation cost of \SYS, we compare the end-to-end cost for all four evaluated workloads on their largest problems.
The time breakdown is shown in \Cref{tab:compile-time}.
We observe that the stage-1 time, including the backtracking and discussing, takes a minor part of the compilation time.
The majority of compile time is in the backend optimizations, which is reasonable since polyhedral schedulers are known to be costly.
However, given larger and more long-running problems than the benchmarks, the compilation will take a smaller fraction of the total time even if the application runs only once, making \SYS practical for real applications.

\begin{table}[h]
    \centering
    \caption{Comparing the compile and running time (in seconds) of \SYS-based programs. The running times listed are at the maximum tested scale.}
    \label{tab:compile-time}
    \begin{tabu}{c|c|
    S[detect-weight, mode=text, table-format=5.2]
    S[detect-weight, mode=text, table-format=5.2]
    S[detect-weight, mode=text, table-format=5.2]
    S[detect-weight, mode=text, table-format=5.2]
    }
        \multicolumn{2}{c|}{\textbf{Stages}}    & \textbf{NPB-BT}   & \textbf{NPB-SP}   & \textbf{NPB-LU}   & \textbf{HPCG} \\
        \hline
                & Stage 1 Execution             & 1.27              & 5.52              & 1.51              & 0.67          \\
        Compile & Stage 2 Optimize              & 280.89            & 1072.42           & 579.42            & 83.49         \\
                & Stage 2 GCC                   & 22.63             & 23.80             & 24.98             & 33.77         \\
        \hline
        \multicolumn{2}{c|}{Run (Stage 2)}      & 10414.35          & 8479.75           & 6753.30           & 539.53        \\
    \end{tabu}
\end{table}

\subsection{Assessing Backend Optimzations}

\subsubsection{Comparing with other polyhedral compilers}

While our targeted implicit solvers with Seidel-style stencils are not covered by existing stencil DSLs, the polyhedral compilers naturally support such computation and can reasonably parallelize them.
We thus compare \SYS with PPCG \cite{ppcg} and PLUTO \cite{pluto} on microbenchmarks to demonstrate the extra effectiveness of our optimization pipeline in such circumstances.
The sequential C++ codes fed into PPCG and PLUTO are generated by the \SYS frontend, without going through our optimizations.
PPCG is invoked with outer-coincidence enforced\footnote{Through the command-line flag \texttt{--isl-schedule-outer-coincidence}.}, which is vital for it to parallelize the test cases with loop-carried dependence.
We enable diamond tiling in PLUTO to get the best performance on it.
We use PLUTO but not PLUTO+ because PLUTO is better maintained and PLUTO+, as an improving fork of PLUTO, focuses on enabling negative coefficients in the linear transformations, which is not useful in our microbenchmarks and will not bring performance improvement.
The microbenchmarks first include symmetric Gauss-Seidel (SymGS) and Incomplete LU (ILU) preconditioners with a 3D 27-point stencil on a $1536^3$ grid.
We also evaluate multiple Jacobi iterations (JacIter) with a 2D 9-point stencil on a $4096^2$ grid, which only involves temporal dependence and is well-optimized by previous stencil optimizations.

\begin{figure}[h]
    \centering
    \includegraphics[width=\linewidth]{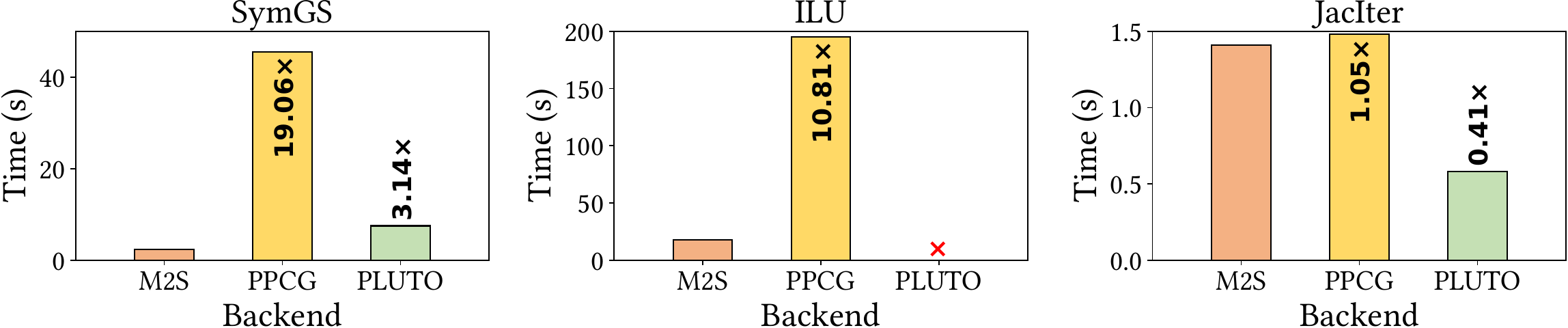}
    \caption{Comparing with PPCG and PLUTO. The performance percentages achieved by PPCG and PLUTO compared to our \SYS are presented alongside. PLUTO times out during compiling ILU. \textbf{Lower is better}.}
    \label{fig:eval-other-poly}
\end{figure}

The result is shown in \Cref{fig:eval-other-poly}.
PPCG consistently performs worse than \SYS, even on Jacobi iterations; it is reasonable since PPCG only performs linear transformations.
PLUTO times out during transpiling the ILU code, taking more than one hour.
Other than that, \SYS outperforms PLUTO by $3.14\times$ on SymGS, demonstrating the effectiveness of our SewTh parallelizer.
Yet, \SYS achieves only $41.4\%$ performance of PLUTO on JacIter, showing that the diamond tiling gives PLUTO outstanding performance on JacIter, but is not available for the Seidel-style stencils.
However, such temporal tiling methods will not help in complicated implicit solvers, since too many loops with their own spatial dependence are attending a single timestep.

\subsubsection{Ablation study on memory optimizations}

\begin{figure}[h]
    \centering
    \includegraphics[width=\textwidth]{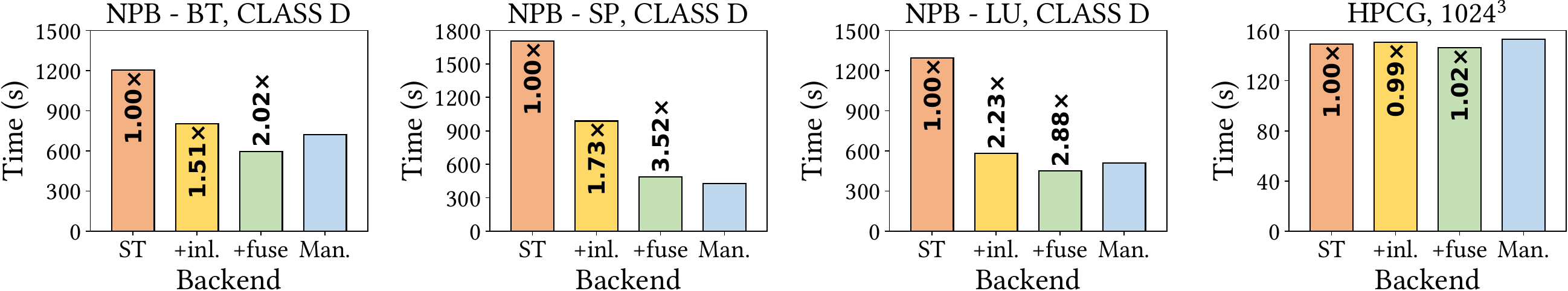}
    \caption{Results of ablation study on memory optimizations. \textbf{ST} is SewTh-only, \textbf{+inl.} adds inlining for elementwise ops, \textbf{+fuse} then adds loop fusion. \textbf{Man.} is manually optimized implementation. \textbf{Lower is better.}}
    \label{fig:eval-ablation}
\end{figure}

To analyze the performance gains of the memory optimizations, we conduct ablation tests on previous benchmarking cases: CLASS D for NPB pseudo-applications, and grid size $1024^3$ for HPCG.
We start from SewTh-only and incrementally introduce inlining and loop fusion.
Together with the manually implemented baseline, the performance numbers are shown in \Cref{fig:eval-ablation}.

It can be seen that both inlining and loop fusion helps a lot in NPB pseudo-applications.
The nonlinearity in their solved Navier-Stokes equation results in a number of elementwise functions and Jacobi-style stencil operators, which will introduce noticeable memory traffic and footprint if not inlined.
Besides, for BT and SP, we observe that the direct solving happens in the three distinct axes, each sweeping forward and backward.
Fusing the forward and backward sweeping loops greatly reduces the size of intermediate tensors by removing two unrelated dimensions, thus significantly improving the performance; the same optimization is manually implemented in the baseline.
Such loop fusion, requiring permuting and reverting the nested loops, is only possible to be automized with a smart enough loop fuser that can select axes and perform affine transforms automatically.

Instead, SewTh-parallelized HPCG already yields good performance matching the manual implementation, and further optimizations help little.
The reason is that HPCG solves a simple Poisson equation through the complicated multigrid conjugate gradient with the SymGS preconditioner and thus \ding{182} requires no elementwise functions, and \ding{183} incurs most of the time cost in SymGS.

\section{Data-Availability Statement}

The implementation of \SYS and baselines corresponded to the evaluation results are available at \cite{artifact} together with the running scripts.
No external dataset is used in the evaluation.

\SYS will also be open-sourced at \url{https://github.com/thu-pacman/Mat2Stencil}.

\section{Related Work}\label{sec:rel}

\paragraph{Solving Differential Equations through Matrix-based Approaches.}

Numerically solving differential equations has been widely treated as sparse linear algebra computation.
Sparse matrix-vector multiplication (SpMV) \cite{spmv2,spmv1,spmv0} are well-studied, covering explicit solving and Jacobi iterations.
Other routines are also available in sparse linear algebra, e.g. SuperLU contains a general-purpose efficient ILU implementation \cite{lishao10}.
SciPy \cite{scipy} is a commonly used Python library for scientific computing and covers many such routines.

While those approaches decouple the equation and solvers, they are too general to be highly optimized against structured grids.
Borrowing their idea of representing differential equation solving in sparse linear algebra, we propose our \SYS based on matrix abstraction, enabling easy algorithm programming while persisting the performance benefits of matrix-free codes.
It is also worth mentioning that \cite{DBLP:conf/pldi/AugustineSP019} takes a totally different approach to optimize a solver already using sparse linear routines, by automatically analyzing the pattern of the sparse matrix and reconstructing matrix-free codes with the best effort.

\paragraph{Solving Differential Equations with Matrix-free Approaches}

The best-studied area of matrix-free approaches is Jacobi-style stencil computations, including with or without time iterations \cite{temporal-stencil,35d-stencil,4d-stencil,Kamil06implicitand}.
Early automatic optimization techniques include \cite{Krishnamoorthy07effectiveautomatic,Datta08stencilcomputation}, which work on manually written nested loops.
In order to provide easier interfaces for programming stencils, researchers have come up with many stencil libraries and DSLs, including \cite{openarray,pochoir,devito,exastencil,pvsc-dtm,snowflake}.
Among those, Devito \cite{devito} and ExaStencils \cite{exastencil} present a set of program representations that is pretty much close to the math formula of the differential equation, being the easiest ones in programming.

However, none of these works fully enables programming implicit solvers, including direct solving of multi-diagonal matrices, (Symmetric) Gauss-Seidel iteration, (Symmetric) Successive Under/Over-Relaxation, Incomplete Lower-Upper \cite{Saad03iterativemethods}, Fine-Grained Parallel Incomplete Lower-Upper \cite{fgpilu}, etc.
The great diversity of such algorithms makes it hard for DSLs to cover all of them. Matrix-free approaches on implicit solvers have been requiring manual optimizations, for example, in \cite{newsw-hpcg} and \cite{gb16}.

\cite{inplace-stencil} introduces automatic code generation for ``in-place stencils'', which is noted as Seidel-style stencils in this paper and helps the construction of implicit solvers, but not includes advanced high-level abstractions required for easier programming of implicit solvers.
Their methodology will also partially degenerate to existing polyhedral approaches for some stencil shapes, e.g. 9 point on 2-D or 27 point on 3-D, in which case our SewTh will perform significantly better.
We cannot compare with their work in evaluation due to absence of their implementation.

\paragraph{Multi-stage Programming}

Our embedded DSL implementation is inspired by Lightweight Modular Staging (LMS) \cite{lms}, which represents multi-staged programming through types in Scala.
We adopt their type-based design to achieve user-transparent staging.
There has also been a Python implementation for it named Snek \cite{snek} that is similar to our staging infrastructure in design, but it does not perfectly feed our needs, thus we implement our own with small tweaks to make the control flow virtualization customizable for different staged types.
Other approaches, e.g. the quote-splice in Scala 3.0 \cite{unified-stage-macro} and lift-annotated MetaML \cite{metaml}, will not fit us well since we do not expect DSL users to get in touch with the staging details.

\paragraph{Tensor Programs}

Our backend generates generalized stencils containing quasi-affine loops that manipulate multi-dimensional arrays, or tensors. Our optimizations originate from various previous works on these types of programs.
Many studies have been taken on optimizing tensor programs since the invention of FORTRAN.
Compiling optimization techniques, especially \textit{polyhedral compilation}, targeting general tensor programs, can be applied automatically \cite{DBLP:books/mk/AllenK2001,DBLP:journals/cacm/PaduaW86,feautrier,pluto,plutoplus,diamond,ppcg}, but the resulting performance is far from optimal for the important Seidel-style stencils.
Nevertheless, compilers implementing such optimizations can be used as a backend to optimize further our solver programs generated by \SYS and we have evaluated against some of them.
Among recent works, \FT \cite{freetensor} is a language and compiler to optimize such a tensor program.
\FT provides a flexible Python interface to construct the tensor program, which is ideal as the target language of our multi-stage programming.
It further provides a set of code-transforming schedules, as performance-tuning primitives.
We use \FT as our backend and our proposed optimizations are implemented as a \FT plugin.

\section{Discussion}\label{sec:dis}

While \SYS has proposed a set of optimizations for the Seidel-style stencils involved in implicit solvers, the optimizations for conventional Jacobi-style stencils are left behind.
It thus leaves huge potential on such optimizations, such as temporal blocking (when possible, it is sometimes forbidden by backward iterations), cache-aware tiling, etc.
Applying such techniques to Seidel-style stencils is a challenging problem left to be solved.

Other than multithreading on CPUs, the latest advancement in accelerators, especially GPGPUs (general-purpose graphics processing units), has gained more and more attention.
The SewTh does not fit into their programming model, which does not allow arbitrary inter-block synchronization.
Algorithmic adjustments are essential as a result, involving techniques like multi-color reordering; incorporating such changes, new compilation techniques need to be developed to automatically parallelize Seidel-style stencils to make them efficiently run on GPGPUs.
Automatic distributed code generation is also possible through polyhedral methods analyzing loop-carried dependences, which has been explored previously \cite{tiramisu}.

Though our current abstraction focuses on Cartesian grids, other types, e.g., star-shaped or hexagonal ones, are possible to be supported with reconsidered predefined matrices.
Speaking even further, semi-structured grids such as block-structured grids might also benefit from our method by combining the unstructured matrix-based representation at the coarse level and our sparse matrix representation for the finer grids, requiring more programming optimization designs.
Besides, our cases only cover the Finite Difference method solvers and new matrices would be required to address Finite Volume and Finite Elements methods.

\section{Conclusion}\label{sec:con}

We propose an innovative domain-specific language (DSL), \SYS, for solving partial differential equations (PDEs) on structured grids.
\SYS introduces a structured sparse matrix abstraction, facilitating modular, flexible, and easy-to-use expression of solvers across a broad spectrum of PDEs, including explicit and implicit solving algorithms.
We evaluate our DSL by implementing four benchmarking programs from the NAS Parallel Benchmarks and High Performance Conjugate Gradients, achieving up to $1.67\times$ and on average $1.03\times$ performance compared to manually matrix-free codes, with significantly less code.

\begin{acks}
  This work is supported in part by NSFC U20B2044 and The Major Key Project of PCL.
\end{acks}

\bibliography{bibfile.bib}



\end{document}